\documentclass[fleqn,usenatbib]{mnras}

\usepackage{newtxtext,newtxmath}

\usepackage[T1]{fontenc}

\DeclareRobustCommand{\VAN}[3]{#2}
\let\VANthebibliography\thebibliography
\def\thebibliography{\DeclareRobustCommand{\VAN}[3]{##3}\VANthebibliography}

\usepackage{graphicx}	
\usepackage{amsmath}	
\usepackage{multirow}
\usepackage{braket}
\usepackage{booktabs}
\usepackage{bm}

\title[Clusters morphology and dynamical state with ZPs]{The Three Hundred Project: quest of clusters of galaxies morphology and dynamical state through Zernike polynomials}

\author[V. Capalbo et al.]
{Valentina Capalbo,$^{1}$\thanks{E-mail: capalbo.1315698@studenti.uniroma1.it}
Marco De Petris,$^{1}$\thanks{E-mail: marco.depetris@uniroma1.it}
Federico De Luca,$^{2,1}$
Weiguang Cui,$^{3}$
\newauthor
Gustavo Yepes,$^{4,5}$
Alexander Knebe,$^{4,5,6}$
Elena Rasia $^{7,8}$
\\
$^{1}$Dipartimento di Fisica, Sapienza Università di Roma, Piazzale Aldo Moro 5, 00185 Roma, Italy\\
$^{2}$Dipartimento di Fisica, Università di Roma “Tor Vergata”, Via della Ricerca Scientifica 1, 00133 Roma, Italy\\
$^{3}$Institute for Astronomy, University of Edinburgh, Royal Observatory, Edinburgh EH9 3HJ, United Kingdom\\
$^{4}$Departamento de Física Teórica, Módulo 8, Facultad de Ciencias, Universidad Autónoma de Madrid, 28049 Madrid, Spain\\
$^{5}$Centro de Investigación Avanzada en Física Fundamental (CIAFF), Facultad de Ciencias, Universidad Autónoma de Madrid, 28049 Madrid, Spain\\
$^{6}$International Centre for Radio Astronomy Research, University of Western Australia, 35 Stirling Highway, Crawley, Western Australia 6009, Australia\\
$^{7}$National Institute for Astrophysics, Astronomical Observatory of  Trieste (INAF-OATs), Via Tiepolo 11, 34131 Trieste, Italy\\
$^{8}$Institute for Fundamental Physics of the Universe (IFPU), Via Beirut 2, 34014 Trieste, Italy
}

\date{Accepted XXX. Received YYY; in original form ZZZ}

\pubyear{2020}

\begin{document}
\label{firstpage}
\pagerange{\pageref{firstpage}--\pageref{lastpage}}
\maketitle

\begin{abstract}
The knowledge of the dynamical state of galaxy clusters allows to alleviate systematics when observational data from these objects are applied in cosmological studies. Evidence of correlation between the state and the morphology of the clusters is well studied.
The morphology can be inferred by images of the surface brightness in the X-ray band and of the thermal component of the Sunyaev-Zel'dovich (tSZ) effect in the millimetre range.
For this purpose, we apply, for the first time, the Zernike polynomial decomposition, a common analytic approach mostly used in adaptive optics to recover aberrated radiation wavefronts at the telescopes pupil plane. With this novel way we expect to correctly infer the morphology of clusters and so possibly, their dynamical state. To verify the reliability of this new approach we use more than 300 synthetic clusters selected in {\small THE THREE HUNDRED} project at different redshifts ranging from 0 up to 1.03. Mock maps of the tSZ, quantified with the Compton parameter, $y$-maps, are modelled with Zernike polynomials inside $R_{500}$, the cluster reference radius. We verify that it is possible to discriminate the morphology of each cluster by estimating the contribution of the different polynomials to the fit of the map. The results of this new method are correlated with those of a previous analysis made on the same catalogue, using two parameters that combine either morphological or dynamical-state probes.
We underline that instrumental angular resolution of the maps has an impact mainly when we extend this approach to high-redshift clusters.
\end{abstract}




\section{Introduction}

Galaxy clusters are the traces of the formation of the largest structures in the Universe and so reliable tools to investigate structures formation and evolution. In principle, this is possible only if and when we have full knowledge of the properties of these objects. The total mass ($i.e.$ the total amount of the Dark Matter (DM), the IntraCluster Medium (ICM) and the stellar components) is an invaluable quantity when exploring the abundances of clusters along the redshift: a standard way to infer cosmological parameters such as the mean matter density $\Omega_m$ and the amplitude of matter perturbations $\sigma_8$ \citep{Planck2016}. Furthermore, under the assumption of a simple self-similar model \citep{Kaiser1986,Voit2005} we could derive the total mass of the clusters from a few observables in optical, X-ray or millimetre band \citep{Giodini2013}. This approach results in a few scaling relations valuable when we are interested to obtain averaged results based on some statistics. However it is prone to the assumed simplified approximations: hydrostatic equilibrium and isothermal and spherical distribution for DM and ICM \citep{Bryan1998}.
It is well known that the hydrostatic equilibrium in haloes is not always satisfied, due to non-thermal pressure contributions from internal motions and turbulence \citep[see e.g.][]{Fang2009,Lau2009,Lagana2010,Rasia2012,Nelson2014,Yu2015,Biffi2016,Eckert2019,Angelinelli2020,Ansarifard2020,Green2020,Gianfagna2020}, pointing out the impact that the dynamical state of those large gravitational bounded objects should have.
Several attempts have been made to infer clusters dynamical state, using both observational data and simulations, by analysing the images of the emission in optical \citep[see e.g.][]{Ribeiro2013,Wen2013} and in the X-ray band \citep[see e.g.][]{Rasia2013,Lovisari2017,Nurgaliev2017,Bartalucci2019,Yuan2020,Cao2020} or of the diffusion of the Cosmic Microwave Background (CMB) photons by tSZ effect in the millimetre band \citep[][hereafter DL20]{Cialone2018,DeLuca2020}, or a combination of some of them \citep[see e.g.][]{Mann2012,Molnar2020,Ricci2020,Zenteno2020,CHEX-MATE2020}.
Among the possibilities we have to mention the studies of the clusters morphology in X-ray and tSZ maps. Several indicators are commonly used, such as: asymmetry parameter \citep{Schade1995}, light concentration \citep{Santos2008}, third-order power ratio \citep{Buote1995,Weissmann2013}, centroid shift \citep{Mohr1993,OHara2006}, strip parameter, Gaussian fit parameter \citep{Cialone2018}, and so on. They exploit the maps with different apertures and efficiencies and are applied individually or combined together, even with different weights \citep[see e.g.][DL20]{Bohringer2010,Nurgaliev2013,Rasia2013,Weissmann2013,Mantz2015,Cui2016,Lovisari2017,Cialone2018,Yuan2020,Cao2020}. 
A complementary approach is by applying thresholds on specific thermodynamic variables. Among the others, the central electron gas density and the core entropy are fairly reliable \citep{Hudson2010}.
The azimuthal scatter in radial profiles of gas density, temperature, entropy or surface brightness \citep{Vazza2011} is also used as a proxy of the ICM inhomogeneities and correlated to the clusters dynamical state \citep[see e.g.][]{Roncarelli2013,Ansarifard2020}.
Alternatively, the projected sky separations between key positions in the images are resulting in reliable estimators of the dynamical state. Interestingly the offsets between the Bright Central Galaxy (BCG) and the peaks and/or the centroids of X-ray or tSZ maps are an indication of how much the relaxation condition is satisfied, with different efficiency \citep[see e.g.][DL20]{Jones1984,Katayama2003,Lin2004,Sanderson2009,Mann2012,Rossetti2016,Lopes2018,Ricci2020,Zenteno2020}.
To be mentioned also other approaches based on wavelets analysis \citep{Pierre1998}, on the Minkowski functionals \citep{Beisbart2001}, or on machine learning \citep[see e.g.][]{Cohn2019,Green2019,Gupta2020}.

In order to explore the morphology of tSZ images of clusters, with the goal to identify features in the signal distribution, for the first time we face the maps with Zernike polynomials \citep[][]{Zernike1934}. Zernike polynomials are largely employed in optical image analysis being an orthogonal basis extended on circular apertures, well fitted to the pupils in optical systems. In the case of adaptive optics, an atmospheric distorted wavefront is modelled with Zernike polynomials in order to generate, with a carefully deformed optical element, a correct wavefront \citep[see e.g.][]{Noll1976,Rigaut1991,Alda1993,Andrade2018}.
In \cite{Ackley2019} these polynomials are used to implement a new automated method to identify astrophysical transients in ground-based surveys.
In the field of image analysis and pattern recognition they are proposed as kernel functions to evaluate image moments with useful invariance properties, such as invariance to image rotation, translation and size \citep[see e.g.][]{Teague1980,Hwang2006,Gao2011}.
Due to the evident capabilities of these polynomials to recover specific patterns hidden in images, many other applications have been proposed, such as in optical metrology or in ophthalmology to describe eye aberrations \citep[see e.g.][]{Liang1997,Thibos2000,Klyce2004,Carvalho2005} and more recent applications in medicine, for example to classify benign and malignant breast masses \citep{Tahmasbi2011}, or to classify organs' shape \citep{Broggio2013} or to study change of osteosarcoma cancer cell lines \citep{Alizadeh2016}.

This innovative approach should be easily applied to large surveys of clusters observed through tSZ signal, to explore their maps within a specific aperture. In this work all the maps are analysed inside a radius equal to $R_{500}$\footnote{$R_{500}$ is the radius of a spherical halo that encloses an overdensity equal to 500 times the critical density of the Universe at a given redshift. A similar definition is used for the subscript 200.}. We use several orders of polynomials, each of them allowing to recover symmetric or asymmetric distributions of ICM at different scales.
The resolution in the maps is a limiting factor to correctly recover the structures. We check as the reliability of this morphological analysis varies in a range of angular resolutions from 5 arcsec to 5 arcmin.

The paper is structured as follows. First we give a basic description of the Zernike polynomials in Sec.~\ref{section:2}. In Sec.~\ref{section:3} the current approaches to derive the clusters morphology from tSZ maps, and in general also from optical and X-ray data, are reviewed. In addition we briefly describe some 3D indicators of the dynamical state available for simulated objects. The dataset used in this work, composed by synthetic clusters generated by state-of-art hydrodynamical simulations, {\small THE THREE HUNDRED} project, is described in Sec.~\ref{section:4}. Finally, the results are shown in Sec.~\ref{section:5} and the consequent conclusions are drawn in Sec.~\ref{section:6}.

\section{Zernike polynomials}
\label{section:2}

The Zernike polynomials are a useful way to analyse images and extract their main features or to model functions on circular aperture. They constitute, in fact, a complete set of orthogonal functions over a unit circle, even with simple invariance properties. 
A common application is in adaptive optics to describe wavefront distortions, more specifically to model the wave aberration function which quantify phase distortions of a wavefront at the exit pupil plane of a telescope. In fact the different polynomials can be also related to classical aberrations of optical systems, such as Gaussian effects of Tilt and Defocus or Seidel ($i.e.$ 3rd order) aberrations of Astigmatism, Coma, Spherical aberration and so on \citep[see e.g.][]{Born1970,Noll1976,Mahajan2006,Lakshminarayanan2011}. In particular it is exploited the property of the Zernike polynomials of representing balanced aberrations, so as to have a wave aberration function decomposed into common shapes of aberrations and with a minimum variance.
For their flexibility in describing inhomogeneities at different scales and azimuthal distributions, we chose these polynomials for fitting maps of tSZ effect inside a circular aperture. A possible alternative are the Hermite-Gauss functions which instead are an orthogonal basis over the whole plane, so more suitable to be applied on rectangular maps.

The Zernike polynomials are defined on a unit circle, therefore it is convenient to describe them with a polar coordinate system so that they result in a product of a radial and an angular function. Following \citet{Noll1976}, they are expressed by:
\begin{equation}
    \label{eq1}
    \begin{cases}
    Z^m_n(\rho,\theta) = N^m_n R^m_n(\rho) \cos{(m\theta)}\\
    Z^{-m}_n(\rho,\theta) = N^m_n R^m_n(\rho) \sin{(m\theta)}
    \end{cases}
\end{equation}
where
\begin{equation}
    \label{eq2}
    N^m_n=\sqrt{\frac{2(n+1)}{1+\delta_{m0}}}
\end{equation}
is a normalization factor in which $\delta_{m0}$ is the Kronecker delta ($\delta_{m0}=0$ if $m\neq0$, $\delta_{m0}=1$ if $m=0$) and
\begin{equation}
    \label{eq3}
    R^m_n(\rho)=\sum_{s=0}^{(n-m)/2} \frac{(-1)^s (n-s)!}{s!\Bigl(\frac{n+m}{2}-s\Bigr)!\Bigl(\frac{n-m}{2}-s\Bigr)!} \rho^{n-2s}
\end{equation}
is the radial term. Each polynomial is a function of $\rho$, the normalised radial distance such that $0\leqslant\rho\leqslant1$ and $\theta$, the azimuthal angle such that $0\leqslant\theta\leqslant2\pi$. The index $n$ defines the order of the radial function and hence of the polynomial, while the index $m$ is the angular frequency. They are positive integers including zero and satisfy $m\leqslant n$ and $n-m=$ even, so that the number of polynomials for a given order $n$ is $n+1$.
Here we follow the ordering method of the different polynomials used in \citet{Noll1976}: starting from the lowest order, $n=0$, for each $n$ we consider at first polynomials with a lower value of $m$. 
\begin{figure}
	\includegraphics[width=\columnwidth]{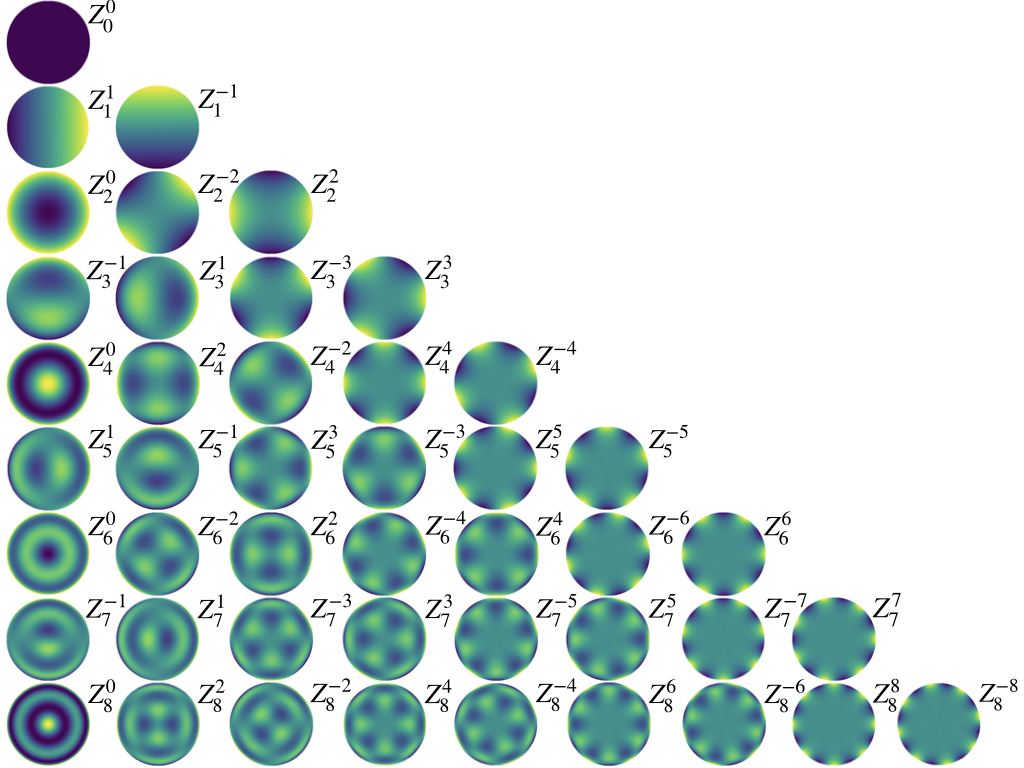}
    \caption{2D maps of 45 Zernike polynomials, $Z_n^{\pm m}$. Each row corresponds to a different polynomial order $n$ and contains $n+1$ terms. From left to right there are polynomials with increasing value of $m$, with $m\leqslant n$ and $n-m=$ even. The colour scale varies from purple to yellow, which respectively indicate the minimum and maximum values of the polynomials.}
    \label{fig:fig1}
\end{figure}
The Fig.~\ref{fig:fig1} shows an ordered list of 45 polynomials used in our analysis \citep[see also][for the mathematical equations]{Mahajan2006}. This scheme easily highlights the difference between the two polynomials of each pair with the same $n$ and $m$ ($i.e.$ $Z^m_n$ and $Z^{-m}_n$): they have the same overall shape, but a different orientation. In this way, any combination of these two paired polynomials will be independent of any angle of rotation with respect to the centre of the circle.
From the 2D maps it is easy to notice axial symmetries and/or antisymmetries of the polynomials. For example, considering the centre of the circle as origin of a reference frame, we can see that polynomials with index $+m$ are symmetric with respect to the horizontal axis whereas polynomials with index $-m$ are antisymmetric with respect to the same axis. In addition, the periodicity of the angular terms implies an invariance of form with respect to rotations of multiples of $2\pi/m$ about the centre. In particular, polynomials with $m=0$ do not have an angular dependence, so that they have a continuous circular symmetry. It is also evident that polynomials with increasing order $n$ are sensitive to smaller spatial scales.

We want to point out that the expressions of Zernike polynomials in Eq.~\eqref{eq1} differ for their normalization from another definition, equally common, used in \citet{Born1970}. However, the conversion between the two expressions is done only by multiplying for a factor $(N_{n}^{m})^{-1}$. Moreover, in literature there are different schemes that use a single index $j$ to order the polynomials sequence, with different conventions \citep[see e.g.][]{Noll1976,Thibos2000}. We prefer to use the double indexing scheme described above, with $n$ and $m$, because it is simple and intuitive since each Zernike term depends upon these two indices. For these reasons, however, care should be taken when comparing results from different works.

Given the above normalization, the polynomials satisfy the following orthogonality property:
\begin{equation}
    \label{eq4}
    \int_{0}^{1}\int_{0}^{2\pi} Z^m_n(\rho,\theta)Z^{m'}_{n'}(\rho,\theta)\rho d\rho d\theta=\pi\delta_{nn'}\delta_{mm'}
\end{equation}
An arbitrary function $\phi(\rho,\theta)$ defined over a unit circular aperture can be expressed as a weighted sum of Zernike polynomials given by:
\begin{equation}
    \label{eq5}
    \phi(\rho,\theta)=\sum_{n=0}^{\infty}\sum_{m=0}^{n} c_{nm}Z_{n}^{m}(\rho,\theta)
\end{equation}
where $c_{nm}$ is a single expansion coefficient (or Zernike moment) resulting from:
\begin{equation}
    \label{eq6}
    c_{nm}=\frac{1}{\pi}\int_{0}^{1}\int_{0}^{2\pi}\phi(\rho,\theta) Z^m_n(\rho,\theta)\rho d\rho d\theta
\end{equation}
The orthogonality of the polynomials ensures that the expansion coefficients do not change when adding further terms. Note that a specific coefficient, $c_{00}$, associated with the polynomial $Z^0_0=1$ ($i.e.$ Piston or Bias) is equal to the mean value of $\phi(\rho,\theta)$ over the unit aperture:
\begin{equation}
    \label{eq7}
    c_{00}=\frac{1}{\pi}\int_{0}^{1}\int_{0}^{2\pi}\phi(\rho,\theta) \rho d\rho d\theta = \braket{\phi}
\end{equation}

\subsection{Fourier transform of Zernike polynomials}
\label{section:2.1}

We have already mentioned that Zernike polynomials are used in different fields as shape descriptors. Taking advantage of their orthogonality, combinations of polynomials can in fact produce a large number of more complex shapes that can fit specific forms with a desired accuracy. Here we applied this approach to study the morphology of galaxy clusters. However, we would like to stress that the goal of our analysis is not to exactly reconstruct the images of clusters by means of polynomials, but rather to highlight only their characteristic features. In particular, we are interested to reveal substructures on large scales, because they are signs of a disturbed dynamical state of the clusters. Our reference spatial scale is of the order of $R_{500}$ for each cluster.
To determine the number of polynomials useful to resolve images on this scale we follow the approach in \citet{Svechnikov2015}, in which the minimum polynomial order $n$ to be used is deduced from simple considerations on the Fourier transform of the polynomials.

Let $F_n^m (k,\phi)$ be the Fourier transform of $Z_n^m (\rho,\theta)$:
\begin{equation}
    \begin{split}
    \label{eq8}
    F_n^m (k,\phi) &= \int\int_{\rho\leqslant 1} Z_n^m (\rho,\theta)\,e^{-2\pi i\,\mathbf{k}\cdot\bm{\rho} } d^2\rho \\ &= \int_{0}^{1}\int_{0}^{2\pi} Z_n^m (\rho,\theta)\,e^{-2\pi i\, k\rho \cos(\theta-\phi) }\rho d\rho d\theta
    \end{split}
\end{equation}
where $\bm{\rho}$ and $\mathbf{k}$ are the position vectors in spatial and frequency domain, respectively, and $\phi$ is the angular coordinate of $\mathbf{k}$.
Using Eq.~\eqref{eq1} for the polynomials, the Fourier transform can be written as:
\begin{equation}
    \begin{cases}
    \label{eq9}
    F_n^m (k,\phi) = (-1)^{n/2+m} \sqrt{\frac{2(n+1)}{1+\delta_{m0}}} \, \frac{J_{n+1}(2\pi k)}{k} \cos(m\phi)\\\\
    F_n^{-m} (k,\phi) = (-1)^{n/2+m} \sqrt{\frac{2(n+1)}{1+\delta_{m0}}} \, \frac{J_{n+1}(2\pi k)}{k} \sin(m\phi)
    \end{cases}
\end{equation}
where $J_{n+1}$ is the Bessel function of the first kind and order $n+1$. In general, a Bessel function $J_n(x)$ oscillates but is not periodic. Its amplitude is maximum when $x\approx n$ and for $x\rightarrow \infty$ decreases asymptotically as $x^{-1/2}$. This means that from Eq.~\eqref{eq9} we can define a limit spatial frequency, $k\approx (n+1)/2\pi$, above which the resolving power ($i.e.$ the power spectrum) of a set of polynomials of order $n$ decays. The spatial frequencies $k$ are in unit of $1/R$, where $R$ is the radius of the circle, therefore in our case they are expressed in unit of $1/R_{500}$. For these reasons, using 45 polynomials shown in Fig.~\ref{fig:fig1}, up to the 8th order, we can be well sensitive to spatial frequencies $k\lesssim 1.4/R_{500}$ in modeling images of galaxy clusters.

An approach similar to that of the Zernike polynomials is the 2D multipole expansion, from which are defined the power ratios \citep{Buote1995}, largely employed in literature for morphological studies of galaxy clusters. In particular is generally used the third-order power ratio, since it is the smallest moment giving an unambiguous account of asymmetric patterns and of the presence of substructures in X-ray or tSZ maps. Here we exploit the possibility to use several polynomials to appreciate symmetric or asymmetric shapes of the ICM and at the same time to distinguish more precisely between circular, elliptical or multi-peaks distributions, defining clearly the resolving scale of the expansion.

\section{Morphological and dynamical parameters}
\label{section:3}

\subsection{2D indicators of morphological regularity}
\label{section:3.1}

A reliable way to infer the dynamical state of galaxy clusters is by studying their morphology, namely the characteristic shapes in 2D projections maps generated by multi-wavelength observations. The first attempts to segregate the clusters between regular and disturbed objects starting from 2D information were applied on X-ray observation maps extracted from Einstein Observatory and ROSAT satellite \citep[see e.g.][]{Jones1992,Mohr1993,Mohr1995,Buote1996,Jones1999}. Afterwards, several parameters have been introduced into the literature and successfully estimated on clusters observed by subsequent X-ray telescopes, such as \textit{Chandra} and XMM-\textit{Newton}, also probing the evolution of dynamical populations with redshift \citep[see e.g.][]{Jeltema2005,Maughan2008,Mann2012,Mantz2015,Parekh2015,Lovisari2017,Bartalucci2019,Yuan2020}.
Hydrodynamical simulations also allowed to validate these parameters with mock X-ray maps \citep[see e.g.][]{Poole2006,Bohringer2010,Morandi2013,Rasia2013,Weissmann2013,Chon2016,Chen2019,Ansarifard2020,Cao2020}.

More recently the morphology has been approached on Sunyaev-Zel'dovich effect observations with a set of indicators similar to those used in X-ray band, at the moment only by simulations \citep[][DL20]{Cialone2018}. Furthermore, to enhance the efficiency of the capability to quantify the morphology, a single parameter is estimated, $M$, both in X-ray and tSZ maps, combining the contributions of a few indicators as follows \citep{Rasia2013,Cialone2018}:
\begin{equation}
    \label{eq10}
    M=\frac{1}{\sum_{i}W_{i}}\Bigg(\sum_{i}W_{i}\frac{\log_{10}(V_{i}^{\alpha_{i}})-\braket{\log_{10}(V_{i}^{\alpha_{i}})}}{\sigma_{\log_{10}(V_{i}^{\alpha_{i}})}}\Bigg)
\end{equation}
here $V_{i}$ denotes the generic $i$th parameter, $W_{i}$ is a weight assigned to each $V_{i}$ related to its efficiency in discriminating the cluster dynamical state, $\alpha_{i}$ is a factor equal to +1 when disturbed clusters are associated with large values of $V_{i}$, otherwise it is equal to -1, the brackets $\braket{\,}$ indicate the average computed over all the clusters and $\sigma$ is the standard deviation.
We use this combined parameter as a reference in describing clusters morphology in tSZ maps to test the efficiency of Zernike polynomials.
$M$ is obtained as a combination of six parameters: the asymmetry parameter ($A$), the light concentration parameter ($c$), the third-order power ratio ($P$), the centroid shift ($w$), the Gaussian fit parameter ($G$) and the strip parameter ($S$) (see \citet{Rasia2013,Cialone2018} for the description of each parameter).
All the details about the estimate of $M$, for the sample of clusters analysed in this work, are reported in DL20.

\subsection{3D indicators of dynamical state}
\label{section:3.2}

In hydrodynamical simulations the dynamical state of each cluster can be derived by the available 3D data. In this way we obtain an \textit{a priori} classification with respect to the morphological one and we can use it to set the best thresholds in 2D parameters to get a good dynamical recognition. Several 3D indicators are used in literature and there is not a unique definition for the criteria and thresholds to discriminate the different clusters states. The classification is also dependent on several factors, including the volume within which the parameters are computed. Furthermore, even the dynamical classes in which clusters can be grouped are not conventionally defined. Details of this kind of analysis are beyond the aim of this paper and we refer to DL20 for a deeper discussion. 

The dynamical state of the clusters of {\small THE THREE HUNDRED} catalogue (see Sec.~\ref{section:4}) was studied in \citet{Cui2018}, \citet{Haggar2020}, DL20, also applying different estimators.
In this work we use the results of the dynamical analysis made in DL20, where the \textit{relaxation parameter}, $\chi$, introduced in \citet{Haggar2020}, is derived with only two 3D indicators.
The first indicator, $f_{s}$, is the ratio of the total substructure masses out of the total cluster mass inside a radius equal to $R_{500}$: $f_{s}=\sum M_{sub}/M_{500}$. The second one, $\Delta_{r}$, is the spatial distance between the position of the centre of mass and of the maximum density peak, normalised to $R_{500}$: $\Delta_{r}=|\mathbf{R}_{cm}-\mathbf{R}_{c}|/R_{500}$. They were used to compute $\chi$ as follow\footnote{Note that the original definition also include the $\eta$ parameter \citep[see also][for detailed definitions of these parameters]{Cui2017}.}:
\begin{equation}
    \label{eq11}
    \chi=\sqrt{ \frac{2}{ \big(\frac{f_s}{0.1}\big)^2 + \big(\frac{\Delta_r}{0.1}\big)^2} }
\end{equation}
Clusters dynamically relaxed are identified with $\chi>1$, while the lower values are related to disturbed systems. In DL20 the correlation between the dynamical state defined with $\chi$ and the morphology quantified with the $M$ parameter is also validated, even along the redshift, allowing us to assume $M$ as a valuable dynamical state discriminator. In particular, with $M<0$ and $M>0$ we recognize, respectively, relaxed and disturbed clusters. Similarly, we check the reliability to infer cluster dynamical state, quantified with $\chi$, by applying Zernike polynomials to recover morphology in tSZ maps.

\section{Synthetic clusters dataset}
\label{section:4}

The proposed approach of studying morphology of galaxy clusters with Zernike polynomials is applied on tSZ maps of synthetic clusters extracted from {\small THE THREE HUNDRED} project\footnote{\url{https://the300-project.org}} \citep{Cui2018}.
The dataset is composed of 324 clusters selected initially among the most massive objects (virial mass $\gtrsim$ 8$\times10^{14}h^{-1}M_\odot$ at $z=0$) in MultiDark simulation \citep{Klypin2016}, $i.e.$ the MultiDark Planck2 simulation. It is generated as a DM-only cosmological box of side length $1h^{-1}$Gpc with $3840^3$ DM particles, each of mass $1.5\times10^9h^{-1}M_{\odot}$, with cosmological parameter values as in \citet{Planck2016} ($h$=0.678, $n$=0.96, $\sigma_8$=0.823, $\Omega_\Lambda$=0.693, $\Omega_m$=0.307 and $\Omega_b$=0.048).
A radius of $15\,h^{-1}$Mpc is used to identify the central region ($i.e$ spherical volumes around the maximum density peak) in the DM haloes that needs to be re-simulated with all the relevant baryonic physics and at higher resolution. The re-simulations are carried out using the SPH code {\small GADGET-X} \citep{Beck2016} and the virialised structures are identified with the AHF (Amiga Halo Finder) \citep{Knollmann2009}. At the end, the obtained sample contains 324 clusters with masses $M_{200}>6\times10^{14}h^{-1}M_{\odot}$ at $z=0$, where $M_{200}$ is the mass inside a radius equal to $R_{200}$.
Here we use all this catalogue in the range of $0\leq z\leq1.03$.

The reliability of the catalogue has been validated studying galaxy properties \citep{Wang2018}, the evolution of the gas density profile \citep{Mostoghiu2019}, the ram pressure stripping on the gas content of haloes and subhaloes \citep{Arthur2019}, the hydrostatic equilibrium mass bias \citep{Ansarifard2020}, backsplash galaxies \citep{Haggar2020}, the filaments in and around cluster environments \citep{Kuchner2020}, alignment of galaxies with respect to the host haloes \citep{Knebe2020}, profiles and distributions of the baryonic components of the clusters \citep{Li2020}, and dynamical state of galaxy clusters from multi-wavelength data (DL20). Other applications are on-going.

\subsection{Maps of thermal Sunyaev-Zel'dovich effect}
\label{section:4.1}

The SZ effect is a distortion in the CMB spectrum due to inverse Compton scattering between photons and energetic free electrons in the ICM \citep{SZ1972,SZ1980}. It can be separated into two components: the kinetic component, which is generated by the proper motion of the cluster with respect to the CMB rest frame and the thermal component, due to random motion of the electrons. Here we analyse only maps of thermal SZ effect. It depends on intrinsic properties of ICM (electron temperature $T_e$ and density $n_e$) and is quantified with the Compton parameter as in the following:
\begin{equation}
    \label{eq12}
    y(\hat{n}) = \frac{\sigma_{T}}{m_e c^2} \int_{los} P_{e}(\hat{n},\ell) d\ell,
\end{equation}
where $P_{e}(\hat{n},\ell)$ is the electron gas pressure (from the ideal gas law $P_e=n_e k_B T_e$, where $k_B$ is the Boltzmann constant) in the direction $\hat{n}$ and along $\ell$, the line of sight ($los$) while $\sigma_{T}$, $m_e$ and $c$ are referring to the Thomson scattering cross-section, the electron mass at rest and the speed of light, respectively.
We emphasize that tSZ maps, $i.e.$ $y$-maps, are sensitive to the diffuse signal of the ICM and distributions of the Compton parameter can extend, in general, over large radii in the clusters. However analysis of these maps, even in internal regions such as within $R_{500}$, show peaks of the signal with low spatial frequencies and extended on quite large scales (see e.g. Fig.~\ref{fig:sections}), so as to be conveniently modelled with low-order Zernike polynomials. Indeed, remember that as their order increases, polynomials become sensitive to finer shapes in images.

Mock $y$-maps of the above clusters catalogue are generated as described in \citet{Cui2018}. The performed procedure consists in discretizing the integration in Eq.~\eqref{eq12}, then obtaining 2D projections of the Compton parameter from a summation extended to all gas particles in a volume $dV$ along the line of sight $dl$:
\begin{equation}
    \label{eq13}
    y=\frac{\sigma_{T}k_{B}}{m_{e}c^{2}dA}\sum_{i}T_{i}N_{e,i}W(r,h_{i})
\end{equation}
where $dA$ is the projected area ($dV=dAdl$), $T_i$ and $N_{e,i}$ are, respectively, the temperature and the number of electrons of the $i$th gas particle, $r$ is the particle position and $h_i$ and $W(r,h_i)$ are the smoothing length and the SPH smoothing kernel adopted in the hydrodynamical simulations. Each $y$-map is derived extending the sum in Eq.~\eqref{eq13} to a distance equal to $\pm1.4 R_{200}$ along the line of sight, being centred on the maximum density peak and with a fixed spatial resolution of 10\,kpc/pixel, in comoving units. Therefore, the pixel size in physical units changes as $1/(1+z)$ at different redshifts. We use 2D projections maps generated along three different directions. We point out that, in this first approach, neither instrumental noise nor astrophysical backgrounds have been considered.

In order to verify the impact of the clusters evolution on the $y$-maps and so on the Zernike polynomials fitting, we take under consideration all the objects sampled at the following three reference redshifts: $z=0$, 0.45 and 1.03. Considering the adopted cosmology, and having fixed the pixel size of each map as described above, the angular resolution is respectively equal to 5.25, 1.14 and 0.59 arcsec at the three redshifts. For each map we only consider an aperture with radius equal to $R_{500}$, to be consistent with the 3D dynamical indicators in DL20. Specifically, we initially place the aperture on the centre of the map, in our case the projected position of the density peak, then we re-centre the aperture on the $y$-centroid within this domain.

\section{Results}
\label{section:5}

\begin{figure*}
    \includegraphics[width=\columnwidth]{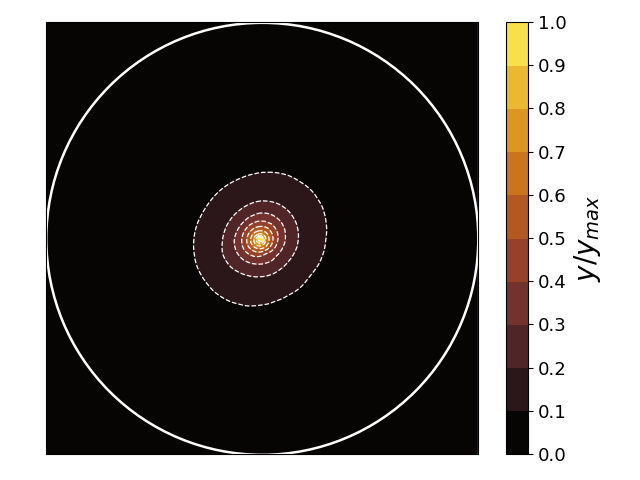}\vspace{5mm}\hfill
    \includegraphics[width=\columnwidth]{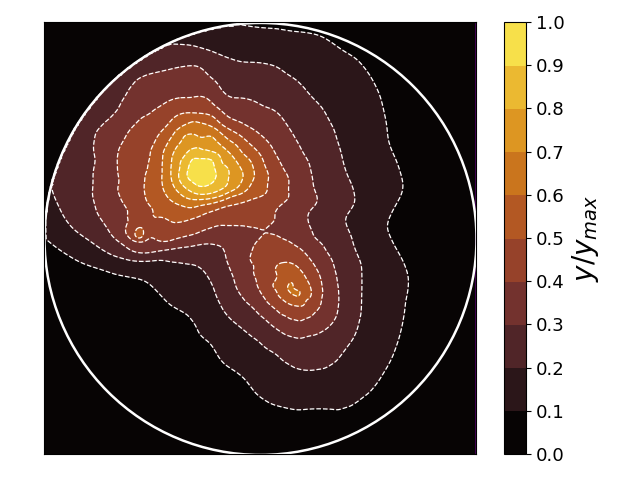}
    \includegraphics[width=\columnwidth]{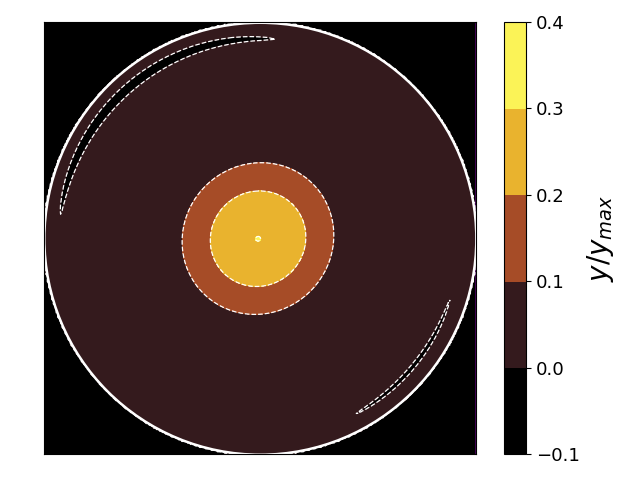}\vspace{5mm}\hfill
    \includegraphics[width=\columnwidth]{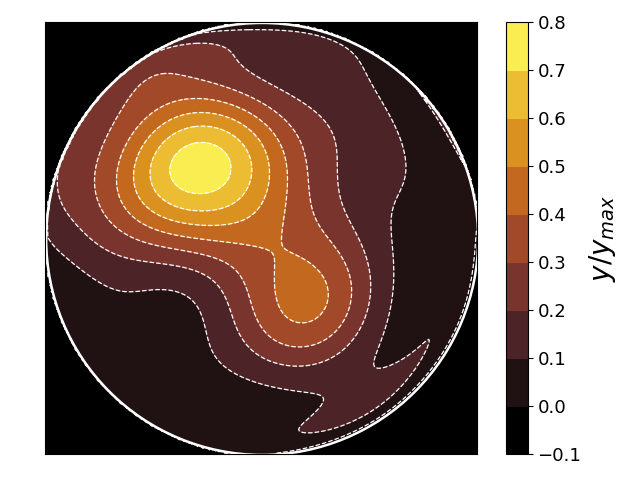}
    \caption{Top panel: $y$-maps of cluster \#299 (left) and cluster \#244 (right) at $z=0$. Bottom panel: Zernike polynomials fitting maps of cluster \#299 (left) and cluster \#244 (right) at $z=0$. The clusters are classified as relaxed (left) and disturbed (right). The isophotes (thin dashed lines) are every 0.1 and the solid circles are the apertures with radius equal to $R_{500}$ ($R_{500}=1.46$\,Mpc for cluster \#299, $R_{500}=1.27$\,Mpc for cluster \#244). The maps are centred on the $y$-centroid and normalized to the maximum value of $y$. The values of $\mathcal{C}$ (see Eq.~\eqref{eq15}) are 0.25 and 3.61 for the cluster \#299 and \#244, respectively.}
    \label{fig:y_maps}
\end{figure*}

The mock $y$-maps of clusters, described in Sec.~\ref{section:4}, are modelled with Zernike polynomials to study their morphology and identify specific features that could help in the clusters dynamical classification.
In Sec.~\ref{section:2.1} we have specified that the frequency resolution that we can obtain with 45 polynomials is $\sim 1.4/R_{500}$, namely the model well fits the $y$-maps up to spatial scales of $\sim 0.7R_{500}$.
From a spectral analysis (see Appendix \ref{appendix:a}) we verify that the Zernike maps remain however able to reveal structures of smaller sizes, up to $\sim 0.5R_{500}$, even if in these cases we cannot reconstruct exactly the signal intensity.
Therefore, we can define a range of spatial frequencies, $k_{min}<k<k_{max}$, well representable by our set of polynomials. Since each $y$-map has a size of $2R_{500}$ in diameter, the minimum frequency is $k_{min}=0.5/R_{500}$, while from the previous considerations the maximum frequency is $k_{max}\sim 2/R_{500}$.
Mainly for these reasons, this method is more suitable for tSZ maps instead of X-ray ones. Indeed, as shown in Sec.~\ref{section:4.1}, the tSZ signal is linearly proportional to the electron density whereas the X-ray emission depends on its second power (the X-ray surface brightness is $S_x\propto\int n_e^2 \Lambda_X dl$, where $\Lambda_X$ is the X-ray cooling function, weakly depending on $T_e$). This means that the tSZ signal is more diffuse than the X-ray one, and the distribution of this latter varies on much smaller spatial scales.

We start our analysis with the first 45 polynomials, up to the order $n=8$ (see Fig.~\ref{fig:fig1}), and for each of them we compute the respective Zernike moment, ${c}_{nm}$. In Appendix \ref{appendix:b} we also analyse the possibility of reducing the number of fitting polynomials without losing sensitivity in reconstructing the morphology. For this analysis we make extensive use of OPTICSPY\footnote{https://github.com/Sterncat/opticspy}, a public Python package for optics applications. The moments result from:
\begin{equation}
    \label{eq14}
    c_{nm}=\frac{\sum y\times Z^m_n}{\pi(R_{500,pixel})^2}
\end{equation}
which is the discrete form of Eq.~\eqref{eq6}, where the sum is extended to all pixels inside the circular aperture and $R_{500,pixel}$ is its radius in pixels. Starting from the lowest orders of polynomials which can model symmetric/asymmetric patterns in horizontal and vertical direction ($i.e.$ $Z_1^{\pm1}$) or circular symmetries ($i.e$ $Z_2^0$), the higher orders allow us to identify small scale image structures.

\begin{figure*}
    \includegraphics[width=\columnwidth]{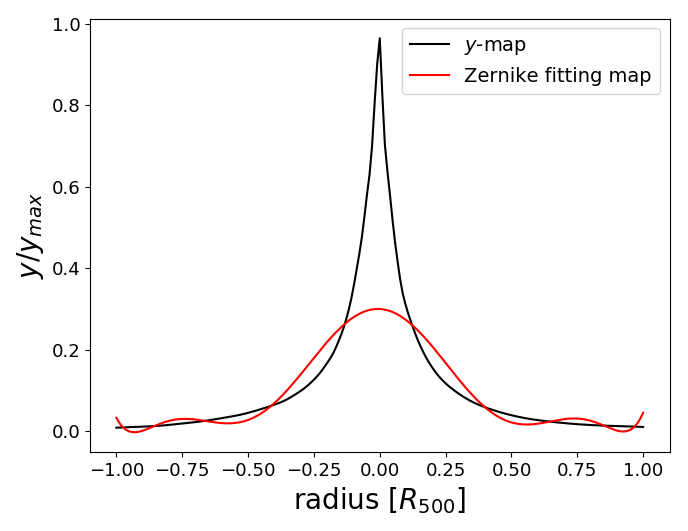}\hfill
    \includegraphics[width=\columnwidth]{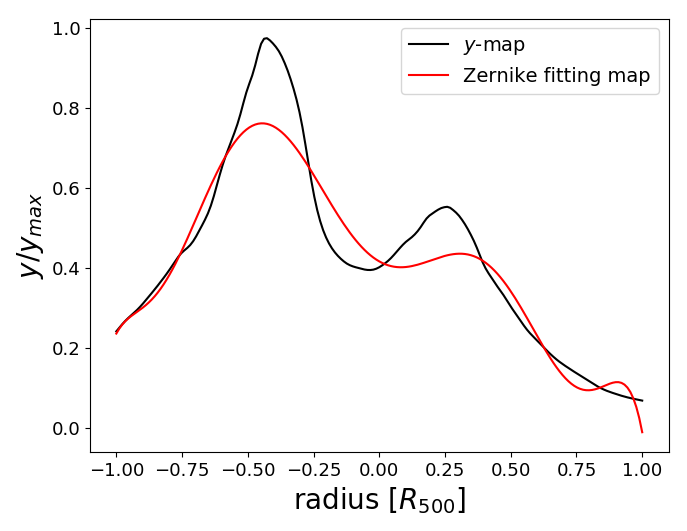}
    \caption{Profiles along a diameter for cluster \#299 (left) and cluster \#244 (right) of its $y$-maps (in black) and Zernike fitting maps (in red), see Fig.~\ref{fig:y_maps}. The two profiles are extracted in vertical direction for the cluster \#299 and oblique direction for the cluster \#244 to intercept the two visible structures.}
    \label{fig:sections}
\end{figure*}

In Fig.~\ref{fig:y_maps} (top panel) the normalised $y$-maps of a relaxed cluster (\#299 in the catalogue) and a disturbed one (\#244) are shown. Their dynamical state is defined by using 3D indicators and through a previous morphological analysis (see Sec.~\ref{section:3}): $\chi=4.06$ and $M=-2.21$ for the relaxed cluster, while $\chi=0.53$ and $M=2.18$ for the disturbed one. We can see that for the relaxed cluster the map has a substantially circular shape, as clearly evident from the isophotes (dashed thin lines). On the contrary, the disturbed cluster has a more complex $y$-map, with an elongated shape and two clear and distinct $y$-peaks indicating a probable on-going merging process. 
In the bottom panel of Fig.~\ref{fig:y_maps} the Zernike polynomials fitting maps of the two clusters above are shown. It is evident that the main features of both images are recovered, even if the signal intensity is not perfectly reconstructed. This is clearly shown in Fig.~\ref{fig:sections}, where profiles along a diameter of the map, extracted both in the $y$-maps and in the fitting ones, are plotted. In the relaxed case the only $y$-peak present is very tight and the polynomial fitting is able to appreciate its symmetry although it cannot exactly reconstruct its amplitude. For the disturbed cluster, on the other hand, the $y$-signal is more smoothed and the polynomial fitting can recognize both peaks.

We verify that the difference between the two fitting maps shown in Fig.~\ref{fig:y_maps} is the contribution of Zernike polynomials with axial symmetries/antisymmetries, $i.e.$ all terms with $m\neq0$. The value of the moments of these polynomials is negligible in the case of relaxed systems. This is what we expect since the relaxed state of a cluster entails the approximation of hydrostatic equilibrium with good isothermality and spherical symmetry, so that the $y$-maps in these cases can easily match polynomials with circular symmetry, $i.e.$ with $m=0$. These terms, however, are present with coefficients of the same order of magnitude also when fitting disturbed clusters. Therefore, to quantify the morphological differences between a relaxed and a disturbed fitting map we can consider the sum of all Zernike moments, in absolute values, related only to polynomials with $m\neq0$. 
For this quantity, from the above considerations, we can expect negligible values in the case of relaxed clusters and larger values in the case of disturbed ones.

\subsection{Zernike polynomials \textit{vs} morphological parameters}
\label{section:5.1}

\begin{figure}
	\includegraphics[width=\columnwidth]{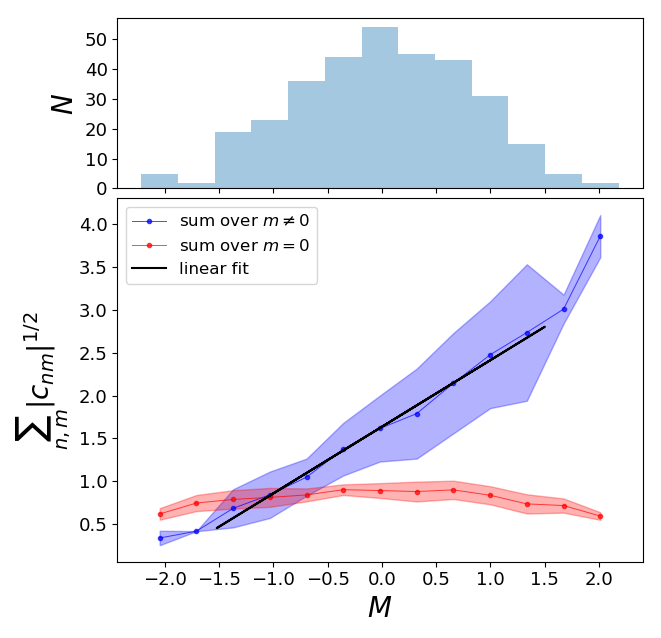}
    \caption{Top: number clusters distribution along the parameter $M$ with binning of 0.3. Bottom: sum over all the fitting Zernike moments $c_{nm}$ with $m\neq0$ (in blue) and $m=0$ (in red) $vs$ the combined parameter $M$, for all clusters at $z=0$. Each point represents the mean value of the sum in each $M$ bin and the colored regions are referring to $\pm1\sigma$. The black line is the fitting line of equation $\mathcal{C}=(0.78\pm0.04)M+(1.64\pm0.03)$ (see Tab.~\ref{tab1}).}
    \label{fig:C_vs_M}
\end{figure}

To validate this new method to recover the clusters morphology, we compare the results from the Zernike fitting with the previous morphological classification made in DL20 on the same catalogue. We use the combined $M$ parameter as comparison tool. We verify that in this case it is convenient to sum the square root of each Zernike moment, in absolute value, with $m\neq0$, so we define:
\begin{equation}
    \label{eq15}
    \mathcal{C}=\sum_{n,\,m\neq0} |c_{nm}|^{1/2}
\end{equation}
In Fig.~\ref{fig:C_vs_M} this sum (in blue) is plotted \textit{vs} $M$, binned in $M$ values, for all clusters catalogue at $z=0$. For comparison, the same sum but extended only over Zernike moments with $m=0$, excluding $c_{00}$ which simply represents the mean value of $y$ over the aperture (see Eq.~\eqref{eq7}), is shown in red. This latter sum seems independent on the morphology being almost constant along $M$. On the contrary, the increasing of the sum in blue is evident throughout the entire range of $M$. Hence, we can discriminate the clusters morphology by computing only $\mathcal{C}$. From this analysis it is clear that the larger value of $\mathcal{C}$ the more disturbed are the clusters. In addition, the width of the colored band shows that the error on each point increases at large $M$, indicating that the images, and then the morphology of the clusters, become more complex, namely more disturbed. Note that, for the invariance properties of the Zernike polynomials summarized in Sec.~\ref{section:2}, $\mathcal{C}$ is invariant for rotations of the maps around the centre.

We estimate a linear proportionality between $\mathcal{C}$ and $M$ expressed by:
\begin{equation}
    \label{eq16}
    \mathcal{C}=a M +b
\end{equation}
The extreme points at maximum and minimum $M$ are prone to the low number of objects in the respective bins, therefore in the linear fit computation we decide to exclude all points in bins with less than 10 clusters. The best-linear fit is shown as a solid black line in Fig.~\ref{fig:C_vs_M}. In Tab.~\ref{tab1} we report the values of the best fit parameters $a$ and $b$ for all the three reference redshifts analysed. The same procedure is repeated considering different angular resolutions, as we discuss later in Sec.~\ref{section:5.3}.
In the last column we also report the Pearson correlation coefficient $r$ between $\mathcal{C}$ and $M$, computed by excluding again the points in the extreme bins.
To estimate the error on the correlation we apply the resampling method of bootstrap at the three redshifts and respective angular resolutions of the catalogue, $i.e.$ $5.25^{\prime\prime}$, $1.14^{\prime\prime}$ and $0.59^{\prime\prime}$. Specifically, we make $10^4$ resamplings of the $\mathcal{C}-M$ dataset and from the distribution of $r$ values we compute the mean and its standard deviation, obtaining: $\braket{r}=0.78\pm0.02$ at $z=0$, $\braket{r}=0.73\pm0.03$ at $z=0.45$ and $\braket{r}=0.75\pm0.02$ at $z=1.03$. Therefore, it is clear that the correlation between the two parameters is strong, and it remains stable at the redshifts we have analysed.

\begin{table}
    \caption{Linear fit parameters $a$ and $b$ (see Eq.~\eqref{eq16}) and Pearson correlation coefficient $r$ between $\mathcal{C}$ and $M$ parameters, for the three different redshifts and the angular resolutions considered in Sec.~\ref{section:5.3}.}
    \centering
    \begin{tabular}{c|c|c|c|c}
    \toprule
    $z$ & Angular resolution & $a$ & $b$ & $r$\\
        & (arcsec)           &     &     &\\
    \midrule
    \multirow{4}*{0} & 5.25 & 0.78 $\pm$ 0.04 & 1.64 $\pm$ 0.03 & 0.78\\
                     & 20   & 0.77 $\pm$ 0.03 & 1.62 $\pm$ 0.03 & 0.78\\
                     & 60   & 0.73 $\pm$ 0.03 & 1.50 $\pm$ 0.02 & 0.79\\
                     & 300  & 0.26 $\pm$ 0.01 & 0.38 $\pm$ 0.01 & 0.77\\
    \midrule
    \multirow{5}*{0.45} & 1.14 & 0.70 $\pm$ 0.04 & 1.63 $\pm$ 0.02 & 0.73\\
                        & 5    & 0.70 $\pm$ 0.04 & 1.60 $\pm$ 0.02 & 0.73\\
                        & 20   & 0.62 $\pm$ 0.03 & 1.29 $\pm$ 0.02 & 0.73\\
                        & 60   & 0.29 $\pm$ 0.02 & 0.42 $\pm$ 0.01 & 0.68\\
                        & 300  & -               & -               & -\\ 
    \midrule                    
    \multirow{5}*{1.03} & 0.59  & 0.68 $\pm$ 0.04 & 1.80 $\pm$ 0.02 & 0.74\\
                        & 5     & 0.65 $\pm$ 0.03 & 1.66 $\pm$ 0.02 & 0.75\\
                        & 20    & 0.36 $\pm$ 0.02 & 0.68 $\pm$ 0.01 & 0.70\\
                        & 60    & -               & -               & -\\
                        & 300   & -               & -               & -\\
    \bottomrule                    
    \end{tabular}
    \label{tab1}
\end{table}

As a second step we also evaluate the correlation between $\mathcal{C}$ and the morphological parameters combined in $M$. In Tab.~\ref{tab2} we report the Spearman rank-order correlation coefficient $r_s$ between $\mathcal{C}$ and each of the six parameters ($V_i$) listed in Sec.~\ref{section:3.1}. We use $r_s$ because we verified that the relationship between $\mathcal{C}$ and the single $V_i$ is not linear. Moreover $r_s$ is less sensitive to the outliers in the tails of the number cluster distributions along $V_i$. We also point out that in DL20 each parameter is estimated inside a proper aperture, while the Zernike polynomials are always estimated inside $R_{500}$.

The correlation of $\mathcal{C}$ with all the parameters, except with the Gaussian fit parameter $G$, is larger than 55\%. With $G$ there is not a clear relationship ($r_s=-0.25$), but this is what we expect since $G$ is not an efficient discriminator \citep[][DL20]{Cialone2018}. It simply measures how much a distribution as a whole differs from a circular symmetry, without taking into account the smaller scale structures.

$\mathcal{C}$ is strongly correlated to the light concentration $c$ ($r_s=-0.85$), one of the best parameters on tSZ maps in discriminating between relaxed and disturbed clusters \citep[][DL20]{Cialone2018}. In principle the two parameters analyse the maps in different ways: $c$ explores the cores in clusters (it was initially introduced to detect cool-core systems, see \citet{Santos2008}) and is suitable for searching very relaxed systems; on the other hand, the Zernike fitting studies the morphology of the maps as a whole and, by combining several polynomials of several orders, it is also able to reveals substructures (as better as the order of the expansion increases, as already discussed). In particular we verify that $\mathcal{C}$ is sensitive to deviations from a pure circular single-peak distribution and therefore it can distinguish very relaxed clusters as does $c$. This can explain the large correlations between these two parameters. 

It is also interesting to analyse the relationship with the centroid shift $w$, which in \citet{Cialone2018} and DL20 is valued as the second
most effective parameter in discriminating the dynamical state of a cluster. $w$ measures how much the centroid varies when it is computed within apertures of different radius, but it can not distinguish between different types of symmetrical distributions around the centre of the aperture. From the above considerations it is evident that the Zernike fitting is not affected by these ambiguities. In particular, we note that $\mathcal{C}$ also differentiates between circular and elliptical patterns: in the first case it is $\sim0$, in the second it has intermediate values around $1.5-2$, and in general it increases if there are substructures.

At last, we remind that the Zernike fitting is a method similar to the computation of the power ratios, which provides a multipole expansion of the signal. However $P$ is just the third-order power ratio ($P=\log_{10}P_3/P_0$), while our analysis combines several polynomials and extends on a larger order of the expansion, then it can distinguish in more detail the various morphologies. In addition we notice that $P$ was computed in DL20 on a smaller aperture (radius equal to $0.5R_{500}$). All these factors can affect the correlation of the two parameters.

By combining multiple functions together, the Zernike fitting becomes sensitive to several aspects of the morphology simultaneously. This explains why $\mathcal{C}$ has a good correlation with different parameters and more generally with their combination $M$.

\begin{table}
    \caption{Spearman rank-order correlation coefficient $r_s$ between $\mathcal{C}$ and each of the six morphological parameters ($V_i$) combined in $M$ (see Sec.~\ref{section:3.1}). The correlation is computed considering all clusters at $z=0$.}
    \centering
    \begin{tabular}{c|c|c|c|}
    \toprule
    Parameter & $r_s$\\
    \midrule
    $A$  & 0.69\\
    $c$  & -0.85\\
    $P$  & 0.56\\
    $w$  & 0.61\\
    $G$  & -0.25\\
    $S$  & 0.61\\
    \bottomrule                    
    \end{tabular}
    \label{tab2}
\end{table}

\subsection{Zernike polynomials \textit{vs} 3D dynamical indicators}
\label{section:5.2}

\begin{figure}
	\includegraphics[width=\columnwidth]{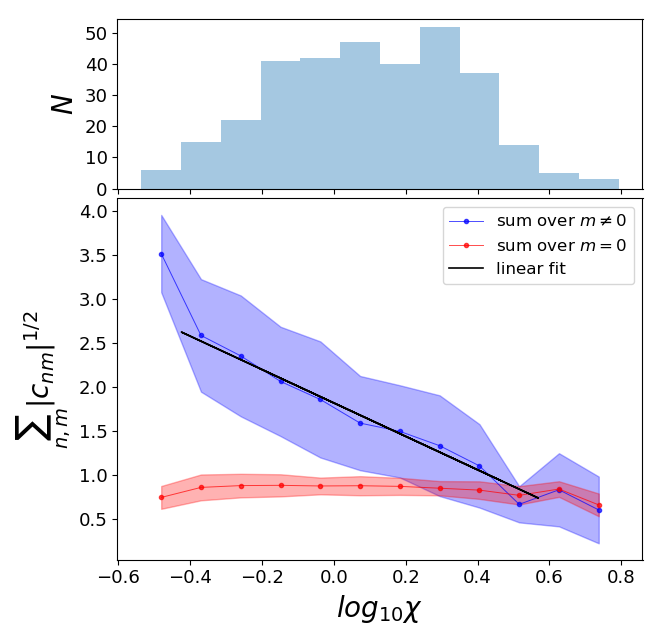}
    \caption{Top: number clusters distribution along $\log_{10}\chi$ with binning of 0.1. Bottom: $\mathcal{C}$ $vs$ $\log_{10}\chi$, for all clusters at $z=0$. Each point represents the mean value of $\mathcal{C}$ in each bin and the colored regions are referring to $\pm1\sigma$. The black line is the fitting line of equation $\mathcal{C}=(-1.90\pm0.14)\log_{10}\chi+(1.82\pm0.04)$ (see Tab.~\ref{tab3}).}
    \label{fig:C_vs_Chi}
\end{figure}

Since in this work we use mock maps of simulated galaxy clusters, we can take advantage of the additional 3D information available to check the goodness of the Zernike analysis in recognize the clusters dynamical state. As done for the comparison with the morphological classification with $M$, we consider the relaxation parameter $\chi$ (see Sec.~\ref{section:3.2}) which provides a rating with continuous values, rather than a discrete classification, of the different dynamical states.

In Fig.~\ref{fig:C_vs_Chi} the $\mathcal{C}$ parameter as defined in Eq.~\eqref{eq15} is plotted, in blue, versus the decimal logarithm of $\chi$. For comparison, in red is plotted the sum of Zernike moments $c_{nm}$ with $m=0$, that has a flat trend as in the previous case. The best-linear fit is shown as solid black line. It is computed neglecting data in bins that have less than 10 clusters. We remind that in this case the relaxed clusters are identified by positive values of $\log_{10}\chi$, so that we obtain the opposite slope with respect to the $M$ parameter:
\begin{equation}
    \label{eq17}
    \mathcal{C}=-a\log_{10}\chi +b
\end{equation}
The best-fitting values of $a$ and $b$ and the Pearson correlation coefficient $r$ are reported in Tab.~\ref{tab3}, for all the redshifts and angular resolutions considered in Sec.~\ref{section:5.3}. 
In Fig.~\ref{fig:C_vs_Chi} is evident a greater dispersion of the points with respect to the linear fit and the error on the fitting slope is also larger than the previous analysis $\mathcal{C}$ $vs$ $M$ (see the fit parameter $a$ in Tab.~\ref{tab1} and Tab.~\ref{tab3}).
The error on the correlation is computed, as in the previous case, by applying the bootstrap resampling at the three redshifts at the respective angular resolutions of the catalogue ($5.25^{\prime\prime}$, $1.14^{\prime\prime}$ and $0.59^{\prime\prime}$). From the distributions of $r$ values we obtain: $\braket{r}=-0.61\pm0.04$ at $z=0$, $\braket{r}=-0.50\pm0.05$ at $z=0.45$ and $\braket{r}=-0.45\pm0.05$ at $z=1.03$. Therefore there is a mild variation of the correlation with redshift. In addition we notice that the correlation between $\mathcal{C}$ and $\chi$ is slightly lower than that between $M$ and $\chi$ reported in DL20 ($\sim66\%$, computed on 10 redshifts between 0 and 1.03). However, we remind that $M$ combines several parameters, while $\mathcal{C}$ is a single indicator, so its application is also satisfying.

We want to stress that the 3D indicators combined in $\chi$ are computed within a spherical aperture centred on the density peak of the clusters, which from an observer's point of view is inaccessible, whereas the $y$-maps are 2D projections of tSZ signal along one direction and we analyse them inside a circular aperture centred on the centroid of $y$. Therefore, in this case it is also interesting to apply the Zernike fitting to maps generated along different lines of sight. Note that all the previous analysis (see Fig.~\ref{fig:C_vs_M} and Fig.~\ref{fig:C_vs_Chi}) are related to maps obtained from projections along the z axis. Now we use all the $y$-maps at $z=0$ also along the x and y axes. We verify that the correlation between $\mathcal{C}$ and $\log_{10}\chi$ is stable around $60\%$ for all the three directions. The respective values of the linear fit parameters $a$, $b$ and the Pearson correlation coefficient $r$ are reported in Tab.~\ref{tab4}. More specifically, the mean value of $\mathcal{C}$ in each bin of $\log_{10}\chi$, for the two directions x and y, varies within $\pm1\sigma$ with respect to the results along z (see Fig.~\ref{fig:C_vs_Chi}). From this analysis it seems clear that the values of $\mathcal{C}$, and the correlation with $\chi$, are not strongly influenced by projection effects.

\begin{table}
    \caption{Linear fit parameters $a$ and $b$ (see Eq.~\eqref{eq17}) and Pearson correlation coefficient $r$ between $\mathcal{C}$ and $\log_{10}\chi$, for the three different redshifts and the angular resolutions considered in Sec.~\ref{section:5.3}.}
    \centering
    \begin{tabular}{c|c|c|c|c}
    \toprule
    $z$ & Angular resolution & $a$ & $b$ & $r$\\
        & (arcsec)           &     &     &\\
    \midrule
    \multirow{4}*{0} & 5.25 & -1.90 $\pm$ 0.14 & 1.82 $\pm$ 0.04 & -0.62\\
                     & 20   & -1.89 $\pm$ 0.14 & 1.80 $\pm$ 0.04 & -0.62\\
                     & 60   & -1.80 $\pm$ 0.13 & 1.67 $\pm$ 0.03 & -0.62\\
                     & 300  & -0.69 $\pm$ 0.05 & 0.45 $\pm$ 0.01 & -0.64\\
    \midrule
    \multirow{5}*{0.45} & 1.14 & -1.47 $\pm$ 0.15 & 1.74 $\pm$ 0.03 & -0.50\\
                        & 5    & -1.47 $\pm$ 0.15 & 1.72 $\pm$ 0.03 & -0.50\\
                        & 20   & -1.39 $\pm$ 0.13 & 1.41 $\pm$ 0.03 & -0.53\\
                        & 60   & -0.70 $\pm$ 0.06 & 0.48 $\pm$ 0.01 & -0.54\\
                        & 300  & -                & -               & -\\ 
    \midrule                    
    \multirow{5}*{1.03} & 0.59 & -1.42 $\pm$ 0.16 & 1.85 $\pm$ 0.03 & -0.45\\
                        & 5    & -1.40 $\pm$ 0.15 & 1.71 $\pm$ 0.03 & -0.47\\
                        & 20   & -0.88 $\pm$ 0.09 & 0.70 $\pm$ 0.02 & -0.50\\
                        & 60   & -                & -               & -\\
                        & 300  & -                & -               & -\\
    \bottomrule                    
    \end{tabular}
    \label{tab3}
\end{table}

\begin{table}
    \caption{Linear fit parameters $a$ and $b$ (see Eq.~\eqref{eq17}) and Pearson correlation coefficient $r$ between $\mathcal{C}$ and $\log_{10}\chi$ parameters for three different directions ($x$, $y$, $z$) along which the $y$-maps are generated, at $z=0$ and with angular resolution of 5.25 arcsec.}
    \centering
    \begin{tabular}{c|c|c|c|}
    \toprule
    Direction & $a$ & $b$ & $r$\\
    \midrule
    x  & -1.84 $\pm$ 0.15 & 1.81 $\pm$ 0.04 & -0.56\\
    y  & -1.83 $\pm$ 0.14 & 1.75 $\pm$ 0.04 & -0.59\\
    z  & -1.90 $\pm$ 0.14 & 1.82 $\pm$ 0.04 & -0.62\\
    \bottomrule                    
    \end{tabular}
    \label{tab4}
\end{table}

\subsection{Impact of angular resolution}
\label{section:5.3}

\begin{figure*}
    \centering
	\includegraphics[scale=0.43]{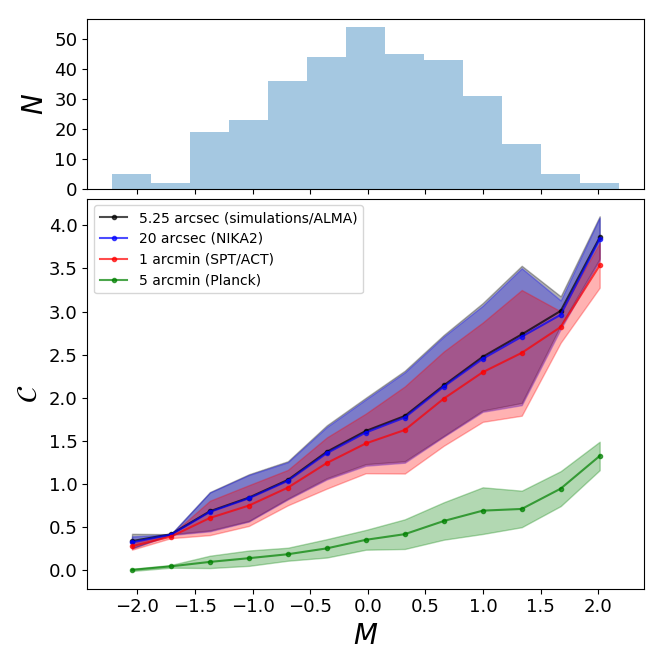}\hspace{1cm}
	\includegraphics[scale=0.43]{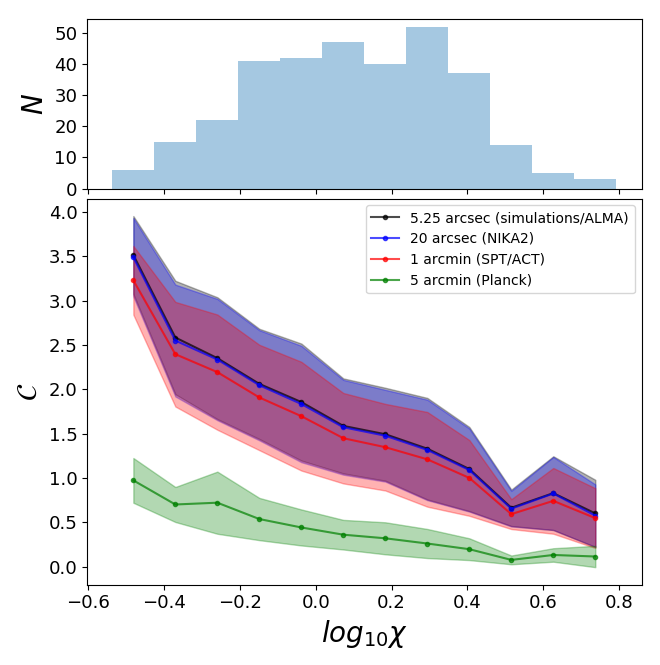}
	\includegraphics[scale=0.43]{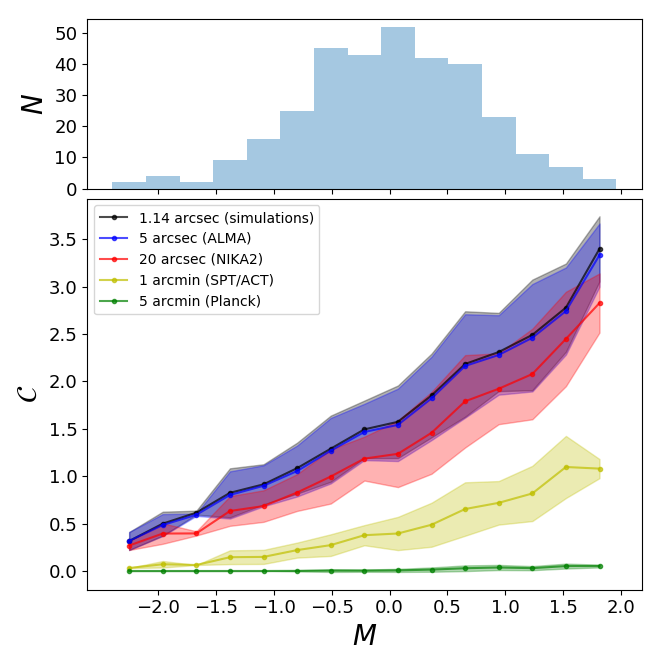}\hspace{1cm}
	\includegraphics[scale=0.43]{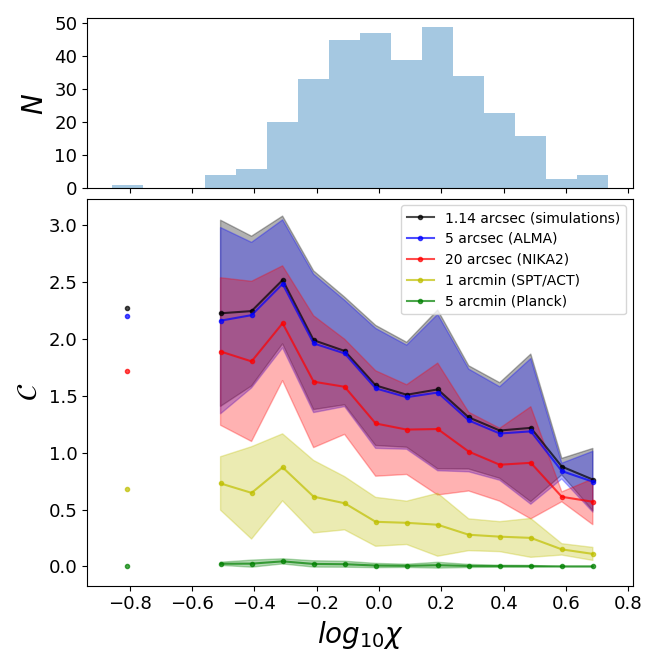}
	\includegraphics[scale=0.43]{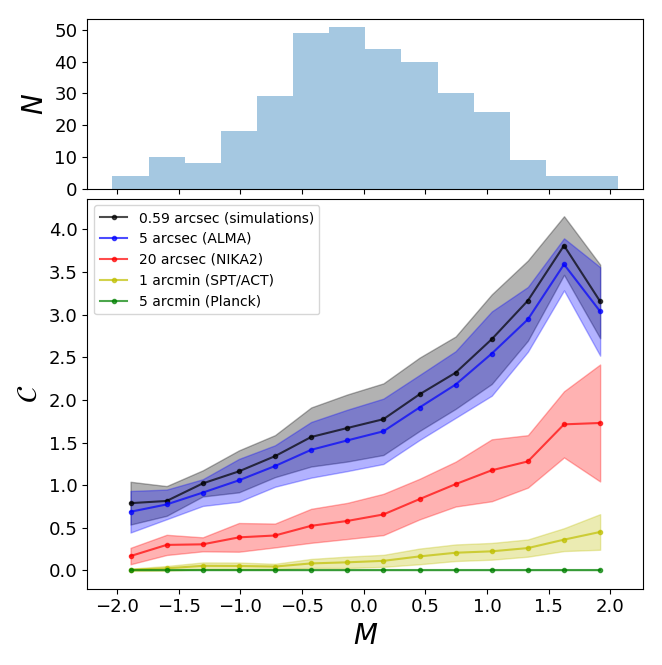}\hspace{1cm}
	\includegraphics[scale=0.43]{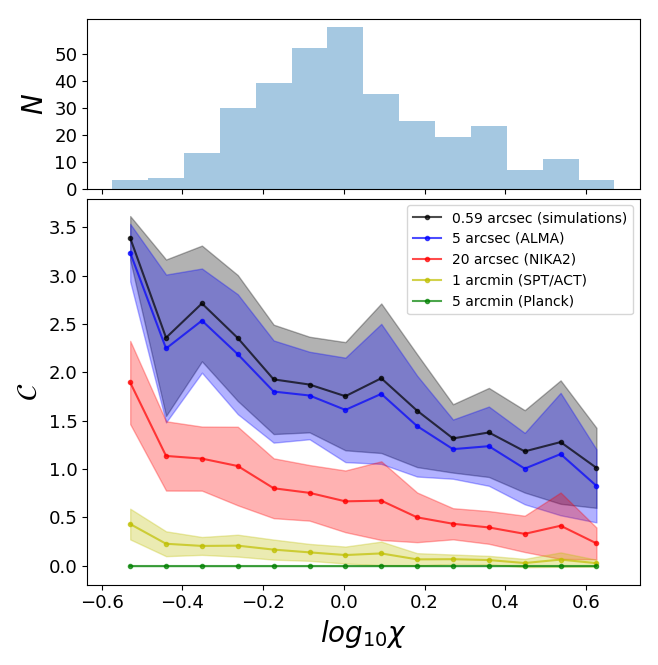}
    \caption{$\mathcal{C}$ $vs$ $M$ (left side) and $\log_{10}\chi$ (right side), binning in $M$ and $\log_{10}\chi$, for all clusters at $z=0$ (top), $z=0.45$ (middle), $z=1.03$ (bottom). Each point represents the mean value of $\mathcal{C}$ in each bin and the respective bands are the regions at $\pm1\sigma$. The different colors are related to the considered resolutions, as reported in the legend. At $z=0$ (top) the black line and shaded area coincide with the blue values. At the top of each panel is shown the distribution of the number of clusters $N$ with respect to $M$ (or $\log_{10}\chi$).}
    \label{fig:smoothing}
\end{figure*}

In real observations the angular resolution of the instrument to generate $y$-maps is a limiting factor to correctly recover the features of the galaxy clusters. For this reason, we examine the impact of different angular resolutions on the proposed approach to infer the morphology. 

We analyse all the clusters catalogue at low, medium and high redshifts ($i.e.$ $z=0$, 0.45 and 1.03), investigating 4 different angular resolutions typical of current millimetre instruments.
We convolve each map with a simple Gaussian beam pattern having several FWHMs corresponding to instruments that have already detected several clusters in millimetric band by means of the tSZ effect. Therefore, we use $5^{\prime\prime}$ as ALMA (Atacama Large Millimeter Array) \citep{Kitayama2016}, $20^{\prime\prime}$ as NIKA2 instrument at IRAM telescope \citep{Adam2018,Perotto2020}, $1^{\prime}$ as SPT (South Pole Telescope) \citep{Bocquet2019,Bleem2020,Huang2020} or ACT (Atacama Cosmology Telescope) \citep{Hasselfield2013,Hilton2018}, and $5^{\prime}$ as \textit{Planck} \citep{Planck2016_SZ}.
In Fig.~\ref{fig:smoothing} we show the $\mathcal{C}$ parameter $vs$ $M$ (left panels) and $\log_{10}\chi$ (right panels), estimated after having convolved the maps with the four angular resolutions, for all the $y$-maps at the three reference redshifts. In each panel the black points are the values for mock $y$-maps, which at $z=0$ have an angular resolution equal to that of ALMA. The linear fit parameters, such as the slope, $a$, the normalization, $b$, as well as the Pearson correlation coefficient, $r$, are reported in Tab.~\ref{tab1} and Tab.~\ref{tab3}. We like remind that points at the extreme limits of $M$ and $\log_{10}\chi$, at all redshifts, are prone to the low number of objects in each bin. We neglect all bins with less than 10 objects. Although at $z=0$ (top panel) even the lowest angular resolution is still able to appreciate the morphology of the disturbed clusters, when we increase the redshift it is clear that we can apply this approach only for high angular resolution (< arcmin) experiments.

\section{Conclusions}
\label{section:6}

The Zernike polynomials, an orthogonal basis of functions defined over unit circular apertures, are largely employed in optical studies. More relevant is their application in adaptive optics, mainly at visible and IR wavelengths, where they describe the wavefront phase distortions on telescope pupil planes induced through atmospheric different phase delay. Due to their capability to recover specific features in images, they have been exploited in other completely different research fields: among the several, in medicine, to analyse the organs' shapes or to map cancer cells distribution patterns in tissues, or in ophthalmology, to describe eye aberrations.
Now, for the first time, we apply these polynomials in clusters of galaxies science to identify their key patterns in maps of the thermal Sunyaev-Zel'dovich (tSZ) effect in order to infer the morphology and, possibly, their dynamical state.
The dynamical classification of clusters, in fact, allows to reduce systematics when these objects are used to infer cosmological parameters.
We use the catalogue of massive clusters of galaxies {\small THE THREE HUNDRED} project, in the redshift range $0\leq z\leq1.03$, as a powerful testbed to validate the capability of this innovative approach.
The analysis is done on mock maps of the Compton parameter, $y$-maps, for each cluster (more than 300, with masses $M_{200}>6\times10^{14}h^{-1}M_{\odot}$ at $z=0$) at three reference redshifts: $z=0$, 0.45 and 1.03. In an independent way, more conventional, the clusters have been segregated by morphology applying the combined parameter $M$, as described in \citet{Cialone2018}, and classified by their dynamical state by 3D indicators available in hydrosimulated objects \citep{DeLuca2020}. Here each $y$-map is modelled with several Zernike polynomials, in order to be able to appreciate shapes of tSZ signal on different spatial scales. More specifically we analyse circular regions with a radius equal to $R_{500}$, centred on the $y$-centroid.
Each polynomial is identified by an index $n$, which defines its degree (or order), and by an index $m$, which determines its angular frequency. We use the first 45 Zernike polynomials, up to the 8th order, and for each of them we compute the respective expansion coefficient (or Zernike moment) $c_{nm}$.

We conclude that a careful choice of only a few of Zernike polynomials results in the capability to recover main features in clusters tSZ images.
In fact, these polynomials are characterized by 2D maps gradually more complex when moving to higher orders $n$, and combined together they can efficiently model the different shapes of the $y$-maps also highlighting the presence of substructures. More precisely, the sum of the absolute value of Zernike moments $c_{nm}$ related to some peculiar polynomials, namely terms with $m\neq0$, is a valuable proxy of the morphological state of a cluster of galaxies. We study the correlation of this quantity with the morphological parameter $M$, a combination of several indicators to better infer the clusters morphology, and the relaxation parameter $\chi$, derived from 3D indicators of the clusters dynamical state. In particular, taking into account only the square root of the Zernike moments, we introduce the parameter $\mathcal{C}=\sum_{n,\,m\neq0} |c_{nm}|^{1/2}$ to correlate, as simply as possible, our results with $M$ and $\chi$. The correlation at $z=0$ is, respectively, $\sim78\%$ and $\sim 62\%$ and it remains fairly stable with redshift in the case of $M$, while for $\chi$ is slightly decreasing.
We would also like to stress that while $M$ combines several parameters, the method we present in this work defines a single estimator, $\mathcal{C}$, which in a similar way is able to appreciate different aspects of the morphology and shows a good performance in discriminating the dynamical state of clusters.

This kind of approach could be easily applied to segregate the population of clusters discovered in large surveys in millimetre band, since it uses a low number of polynomials.
For this purpose we have also analysed the possibility of reducing the polynomials used to a lower number, which in any case remains efficient in discriminating the morphology, in order to speed up the computational time.

We verify that the method is, as expected, prone to the final angular resolution of the maps of the clusters. 
The morphology of high-redshifts clusters ($z\simeq1$) could be recovered only with instruments with sub-arcminute resolution.

This analysis is also extended to X-ray maps, and the first application is on the $y$-maps produced by the \textit{Planck} satellite, starting with the local clusters in the catalogue. These are the topics of works in preparation.

\section*{Acknowledgements}
We thank F. Pedichini, M. Stangalini and A. Terreri for useful comments on wavefront modeling on telescope pupils in the first phase of this work. 

This work has been made possible by {\small THE THREE HUNDRED} collaboration. The simulations used in this paper have been performed in the MareNostrum Supercomputer at the Barcelona Supercomputing Center, thanks to CPU time granted by the Red Espa\~nola de Supercomputaci\'on. As part of {\small THE THREE HUNDRED} project, this work has received financial support from the European Union's Horizon 2020 Research and Innovation programme under the Marie Sklodowskaw-Curie grant agreement number 734374, the LACEGAL project.

VC, MDP and FDL acknowledge support from Sapienza Università di Roma thanks to Progetti di Ricerca Medi 2019, prot. RM11916B7540DD8D. WC acknowledges support from the European Research Council under grant number 670193. GY and AK acknowledge financial support from \textit{Ministerio de Ciencia, Innovación y Universidades / Fondo Europeo de DEsarrollo Regional}, under research grant PGC2018-094975-C21.
AK further acknowledges support from the Spanish Red Consolider MultiDark FPA2017-90566-REDC and thanks Nick Corbin for the sweet escape.
ER acknowledges funding under the agreement ASI-INAF N.2017-14-H.0.

\section*{Data Availability}
The data used in this paper are part of {\small THE THREE HUNDRED} project and can be accessed following the guide lines of the collaboration that can be found in the website \url{https://the300-project.org}. The data specifically shown in this paper will be shared upon request to the authors.




\bibliographystyle{mnras}
\bibliography{biblio} 


\appendix

\section{Power spectra}
\label{appendix:a}

\begin{figure}
    \includegraphics[width=\columnwidth]{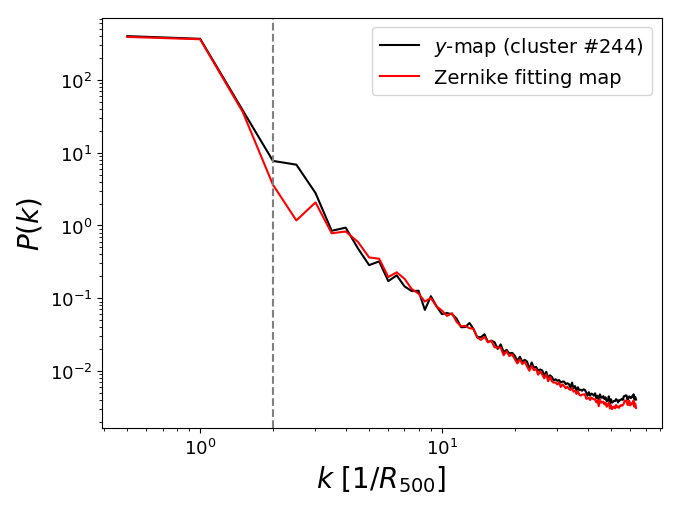}\vspace{5mm}
    \includegraphics[width=\columnwidth]{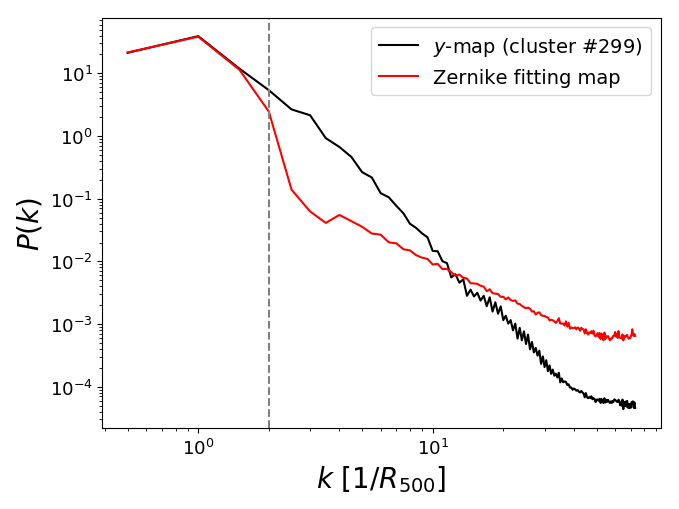}
    \caption{Power spectrum for cluster \#244 (top) and cluster \#299 (bottom) $vs$ spatial frequency $k$ in unit of $1/R_{500}$. The black lines refer to the $y$-maps and the red ones to the Zernike fitting maps. The vertical shaded lines in grey correspond to $k=2/R_{500}$, the maximum spatial frequency appreciable by the Zernike fitting, according to the considerations reported in the text below.}
    \label{fig:power_spectra}
\end{figure}

To define the spatial scale at which the Zernike fitting is sensitive we compare the power spectra of the $y$-maps and of the respective Zernike fitting maps. In Fig.~\ref{fig:power_spectra} we show two examples of the same clusters analysed in Sec.~\ref{section:5}, namely the disturbed cluster \#244 (top) and the relaxed cluster \#299 (bottom). The power spectra are derived from \citep{Svechnikov2015}:
\begin{equation}
    \label{eqA1}
    P(k)=\frac{1}{S}\frac{1}{2\pi} \int_{0}^{2\pi} |F(k,\phi)|^2\,d\phi
\end{equation}
where $S$ is the area of the circular aperture and $F(k,\phi)$ is the Fourier transform of the image. We compute the spectra by discretizing this equation, that is by adding azimuthally all the pixels at a given frequency $k$. In both cases the spectrum of the Zernike map overlaps that of the $y$-map for $k\lesssim1.4/R_{500}$, as expected from considerations made in Sec.~\ref{section:2.1}, and it remains able in modeling the slope for the disturbed cluster also at higher frequencies.
This good matching is generally showed for maps with beveled shapes and quite large peaks (in terms of spatial scales), which results in power spectra that quickly collapse at high spatial frequencies. In particular, we have extracted a sample of very disturbed clusters (8 objects with $\log_{10}\chi<-0.4$, $M>0.5$ and $\mathcal{C}>2.5$ at $z=0$) and we have verified that their spectra all fall in this category.
In the relaxed case, on the contrary, the spectrum of the $y$-map drops more slowly and the fit gets worse for $k\gtrsim2/R_{500}$.
We have also extracted a sample of highly relaxed clusters (6 objects with $\log_{10}\chi>0.6$, $M<-0.5$ and $\mathcal{C}<1.5$ at $z=0$), and we have verified that for $k>2/R_{500}$ in all cases there is a more or less drastic decrease of the power spectrum of the polynomials. For this reasons we consider the frequency $k\sim2/R_{500}$ as limit value of the resolving scale of our Zernike fitting.

\section{Contribution of the different Zernike polynomials}
\label{appendix:b}

\begin{figure}
	\includegraphics[width=\columnwidth]{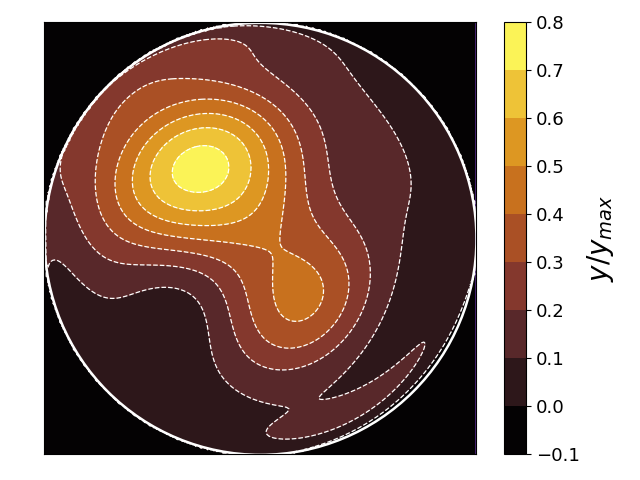}
    \caption{Zernike fitting map of cluster \#244 obtained with a reduced number of polynomials. We considered all terms with $0\leq n\leq8$ and $0\leq m\leq3$, namely 27 polynomials instead of 45 as in Fig.~\ref{fig:y_maps} (bottom right panel).}
    \label{fig:244_reduced}
\end{figure}

\begin{figure}
	\includegraphics[width=\columnwidth]{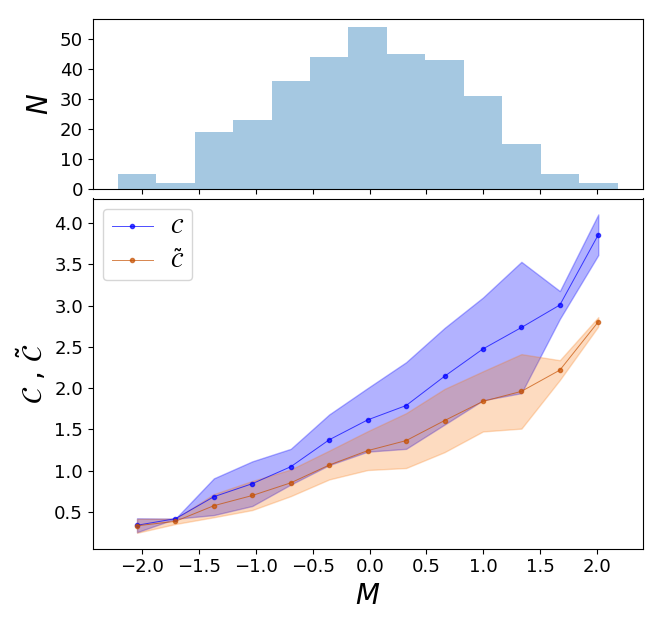}\vspace{5mm}
	\includegraphics[width=\columnwidth]{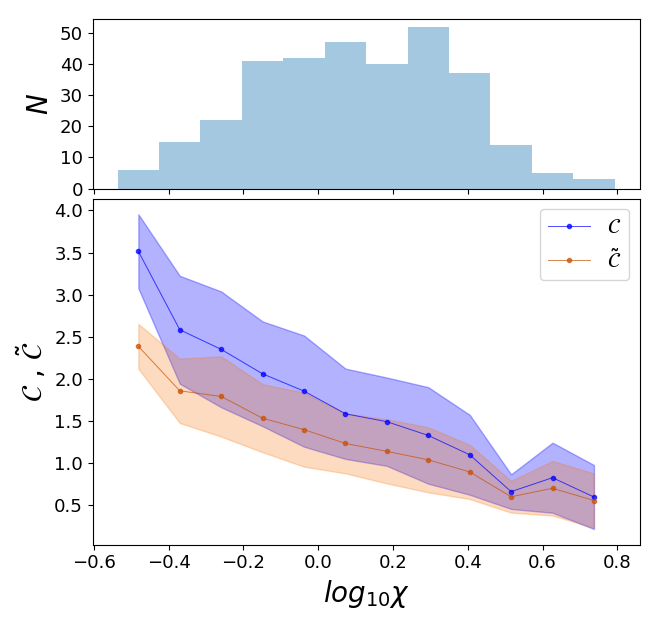}
    \caption{$\mathcal{C}$ (in blue) and $\Tilde{\mathcal{C}}$ (in orange) $vs$ $M$ (top) and $\log_{10}\chi$ (bottom), binning in $M$ and $\log_{10}\chi$ values respectively, for all clusters at $z=0$. Each point is the mean value of $\mathcal{C}$ (or $\Tilde{\mathcal{C}}$) in each bin and the respective bands are the regions at $\pm1\sigma$.}
    \label{fig:C_reduced}
\end{figure}

\begin{figure*}
    \includegraphics[scale=0.44]{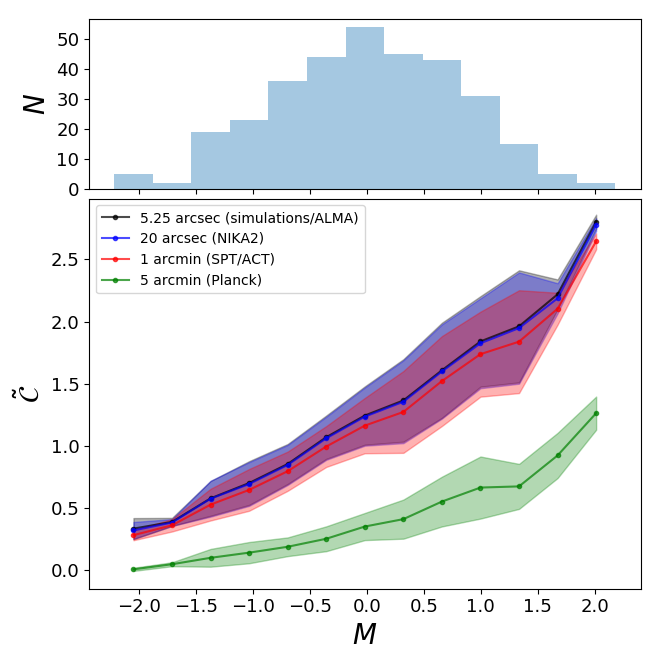}\hspace{1cm}
	\includegraphics[scale=0.44]{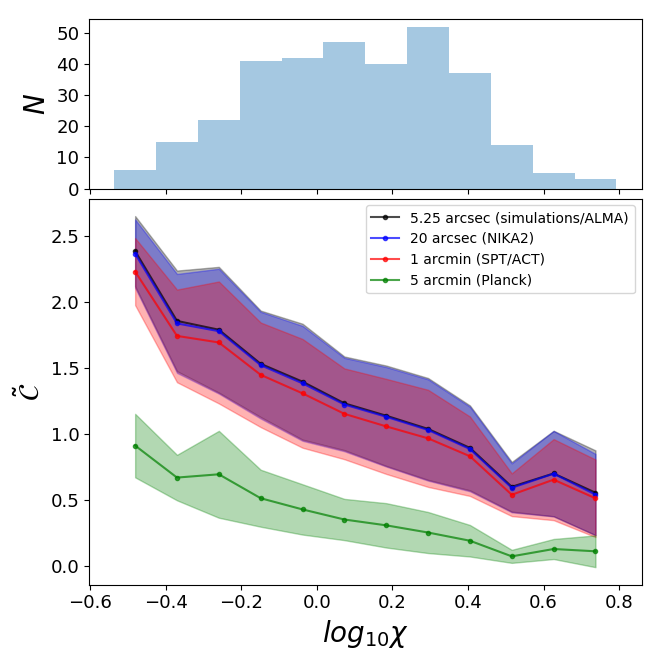}
	\includegraphics[scale=0.44]{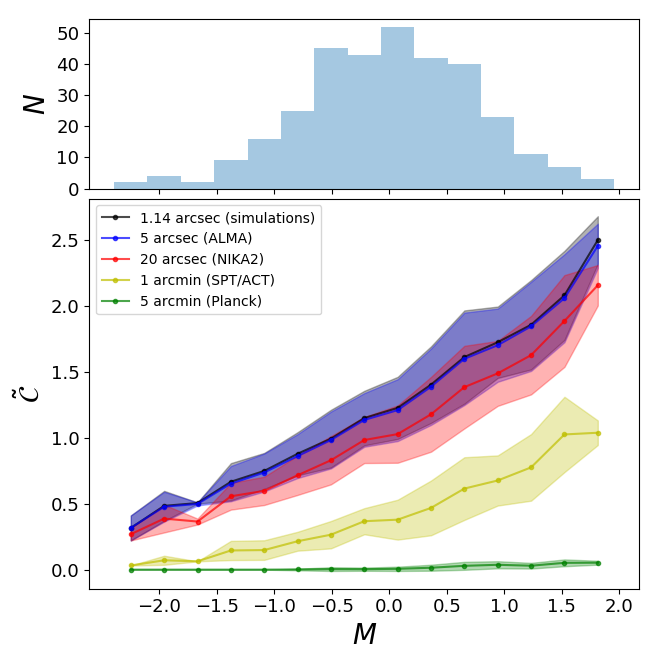}\hspace{1cm}
	\includegraphics[scale=0.44]{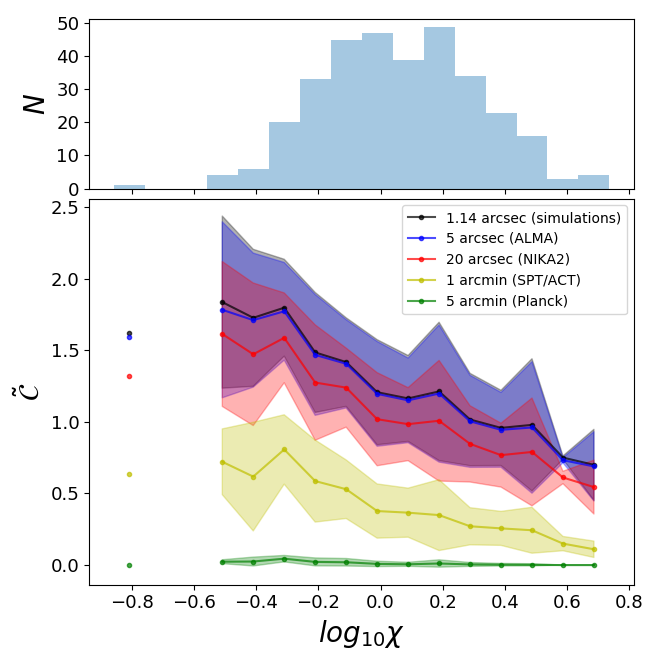}
	\includegraphics[scale=0.44]{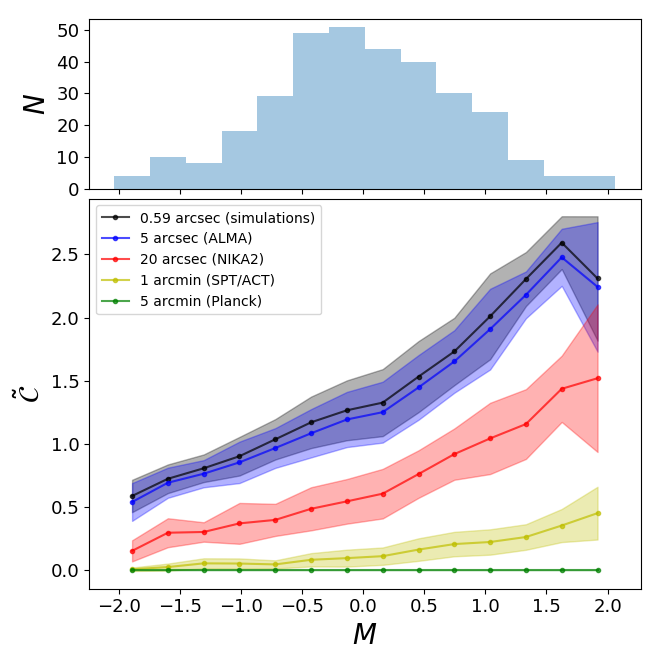}\hspace{1cm}
	\includegraphics[scale=0.44]{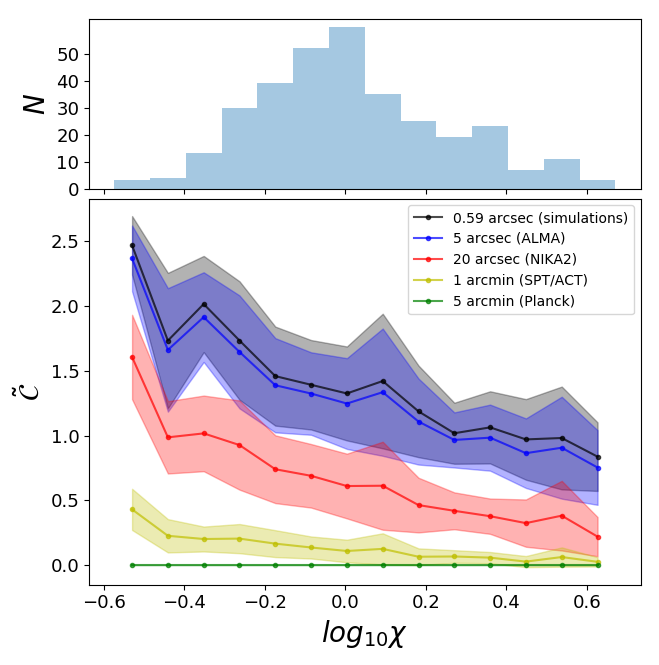}
    \caption{$\Tilde{\mathcal{C}}$ $vs$ $M$ (left side) and $\log_{10}\chi$ (right side), binning in $M$ and $\log_{10}\chi$, for all clusters at $z=0$ (top), $z=0.45$ (middle), $z=1.03$ (bottom). Each point represents the mean value of $\Tilde{\mathcal{C}}$ in each bin and the respective bands are the regions at $\pm1\sigma$. The different colors are related to the considered resolutions, as reported in the legend. At the top of each panel is shown the distribution of the number of clusters $N$ with respect to $M$ (or $\log_{10}\chi$).}
    \label{fig:smoothing_reduced}
\end{figure*}

We started our study of the clusters morphology using 45 Zernike polynomials, up to the order $n=8$, in order to best model the various forms of the signal that may occur. Nevertheless, we verify that polynomials with high values of azimuthal frequency $m$, although not null in the case of disturbed clusters, give a negligible contribution in reconstructing the morphology of the Compton parameter for our catalogue.
This is clearly due to the nature of the tSZ signal, which is quite smooth. The value of the expansion coefficients of the polynomials with $m>3$ is small, so that they do not improve the dynamical class segregation capability. Hence, we reduce the number of terms to use without losing the morphological information we need. For the values of $m$ between 0 and 3, however, we consider all the respective orders $n$ between those previously used. In this way, in fact, the polynomial fitting remains sensitive to the signal on smaller scales, as shown in Fig.~\ref{fig:244_reduced}, in which the elongated pattern in direction of the two substructures of cluster \#244 is still well evident. The two Zernike fitting maps, obtained respectively with 45 and 27 polynomials, have a pixel to pixel difference less than 0.06, and the rms of their difference is 0.016. All the polynomials excluded from the analysis have $m\neq0$. For this reason we only show the fitting map of a disturbed cluster because, as previously described, these terms do not impact the results for a relaxed cluster.
From these considerations we can consider a reduced form of $\mathcal{C}$ to correlate the Zernike fitting results with $M$ and $\log_{10}\chi$, namely:
\begin{equation}
    \label{eqB1}
    \Tilde{\mathcal{C}}=\sum_{\substack{n \\ 0 < m\leq3}} |c_{nm}|^{1/2}
\end{equation}
In Fig.~\ref{fig:C_reduced} we compare $\mathcal{C}$ and $\Tilde{\mathcal{C}}$ $vs$ $M$ (top panel) and $\log_{10}\chi$ (bottom panel), for all clusters catalogue at $z=0$. It is evident that even the reduced parameter $\Tilde{\mathcal{C}}$ reproduces well the previous trends and it remains able in discriminating the dynamical state of galaxy clusters. The equations \eqref{eq16} and \eqref{eq17} are still also valid and all the fit parameters concerning the slopes of $\Tilde{\mathcal{C}}$ $vs$ $M$ and $\Tilde{\mathcal{C}}$ $vs$ $\log_{10}\chi$ are reported in Tab.~\ref{tabB1} and Tab.~\ref{tabB2}. We neglect all bins in $M$ and $\log_{10}\chi$ with less than 10 clusters.
The correlation with $M$ and $\log_{10}\chi$ is similar to the previous case, for all the three reference redshifts considered, and the value of $\Tilde{\mathcal{C}}$ is $\sim30\%$ lower than $\mathcal{C}$. A reduced number of polynomials could be useful in case of very large catalogues, to reduce the computational time of the fitting routine. Therefore, for completeness, we show in Fig.~\ref{fig:smoothing_reduced} the results of the same analysis described in Sec.~\ref{section:5.3} but considering the reduced parameter $\Tilde{\mathcal{C}}$. The parameters derived from the respective linear fits, namely the slope $a$, the normalization $b$, and the Pearson correlation coefficient $r$, are reported in Tab.~\ref{tabB1} and Tab.~\ref{tabB2}.

\begin{table}
    \caption{Linear fit parameters $a$ and $b$ and Pearson correlation coefficient $r$ between $\Tilde{\mathcal{C}}$ and $M$, for the three different redshifts and the angular resolutions considered in Sec.~\ref{section:5.3}.}
    \centering
    \begin{tabular}{c|c|c|c|c}
    \toprule
    $z$ & Angular resolution & $a$ & $b$ & $r$\\
        & (arcsec)           &     &     &\\
    \midrule
    \multirow{4}*{0} & 5.25 & 0.54 $\pm$ 0.02 & 1.25 $\pm$ 0.02 & 0.82\\
                     & 20   & 0.53 $\pm$ 0.02 & 1.24 $\pm$ 0.02 & 0.82\\
                     & 60   & 0.51 $\pm$ 0.02 & 1.17 $\pm$ 0.02 & 0.82\\
                     & 300  & 0.25 $\pm$ 0.01 & 0.37 $\pm$ 0.01 & 0.77\\
    \midrule
    \multirow{5}*{0.45} & 1.14 & 0.49 $\pm$ 0.02 & 1.24 $\pm$ 0.01 & 0.76\\
                        & 5    & 0.49 $\pm$ 0.02 & 1.23 $\pm$ 0.01 & 0.76\\
                        & 20   & 0.44 $\pm$ 0.02 & 1.05 $\pm$ 0.01 & 0.76\\
                        & 60   & 0.27 $\pm$ 0.02 & 0.41 $\pm$ 0.01 & 0.71\\
                        & 300  & -               & -               & -\\ 
    \midrule                    
    \multirow{5}*{1.03} & 0.59 & 0.48 $\pm$ 0.02 & 1.35 $\pm$ 0.01 & 0.77\\
                        & 5    & 0.46 $\pm$ 0.02 & 1.28 $\pm$ 0.01 & 0.78\\
                        & 20   & 0.31 $\pm$ 0.02 & 0.63 $\pm$ 0.01 & 0.72\\
                        & 60   & -               & -               & -\\
                        & 300  & -               & -               & -\\
    \bottomrule                    
    \end{tabular}
    \label{tabB1}
\end{table}

\begin{table}
    \caption{Linear fit parameters $a$ and $b$ and Pearson correlation coefficient $r$ between $\Tilde{\mathcal{C}}$ and $\log_{10}\chi$, for the three different redshifts and the angular resolutions considered in Sec.~\ref{section:5.3}.}
    \centering
    \begin{tabular}{c|c|c|c|c}
    \toprule
    $z$ & Angular resolution & $a$ & $b$ & $r$\\
        & (arcsec)           &     &     &\\
    \midrule
    \multirow{4}*{0} & 5.25 & -1.29 $\pm$ 0.09 & 1.38 $\pm$ 0.02 & -0.62\\
                     & 20   & -1.28 $\pm$ 0.09 & 1.37 $\pm$ 0.02 & -0.62\\
                     & 60   & -1.24 $\pm$ 0.09 & 1.29 $\pm$ 0.02 & -0.63\\
                     & 300  & -0.66 $\pm$ 0.04 & 0.44 $\pm$ 0.01 & -0.65\\
    \midrule
    \multirow{5}*{0.45} & 1.14 & -0.95 $\pm$ 0.10 & 1.32 $\pm$ 0.02 & -0.48\\
                        & 5    & -0.95 $\pm$ 0.10 & 1.30 $\pm$ 0.02 & -0.48\\
                        & 20   & -0.93 $\pm$ 0.09 & 1.12 $\pm$ 0.02 & -0.52\\
                        & 60   & -0.65 $\pm$ 0.06 & 0.46 $\pm$ 0.01 & -0.55\\
                        & 300  & -                & -               & -\\ 
    \midrule                    
    \multirow{5}*{1.03} & 0.59 & -1.01 $\pm$ 0.11 & 1.38 $\pm$ 0.02 & -0.47\\
                        & 5    & -1.00 $\pm$ 0.10 & 1.31 $\pm$ 0.02 & -0.49\\
                        & 20   & -0.78 $\pm$ 0.07 & 0.65 $\pm$ 0.02 & -0.53\\
                        & 60   & -                & -               & -\\
                        & 300  & -                & -               & -\\
    \bottomrule                    
    \end{tabular}
    \label{tabB2}
\end{table}

\bsp	
\label{lastpage}
\end{document}